\pdfoutput=1
\newif\ifarxiv
\arxivtrue
\pdfmapfile{+classico.map}
\newif\ifafour
\afourfalse% true = A4, false = A5
\newif\iftypodisclaim % typographical disclaim on the side
\typodisclaimtrue

\documentclass[\ifafour a4paper,12pt,\else a5paper,10pt,\fi%extrafontsizes,%
onecolumn,oneside,article,%french,italian,german,swedish,latin,
british%
]{memoir}
\newcommand*{\firstpublished}{13 December 2019}
\newcommand*{\updated}{\ifarxiv 6 May 2023\else\today\fi}
\newcommand*{\propertitle}{Dimensional analysis in relativity\\ and in differential geometry%\\{\large ***}%
}% title uses LARGE; set Large for smaller
\newcommand*{\pdftitle}{Dimensional analysis in relativity and in differential geometry}
\newcommand*{\headtitle}{Dimensional analysis, relativity, manifolds}
\newcommand*{\pdfauthor}{P.G.L.  Porta Mana}
\newcommand*{\headauthor}{Porta Mana}
\newcommand*{\reporthead}{\ifarxiv\else Open Science Framework \href{https://doi.org/10.31219/osf.io/jmqnu}{\textsc{doi}:10.31219/osf.io/jmqnu}\fi}% Report number

\usepackage[T1]{fontenc} 
\input{glyphtounicode} \pdfgentounicode=1

\usepackage[utf8]{inputenx}

\usepackage{textcomp}

\usepackage{amsmath}

\usepackage{mathtools}
\usepackage{amssymb}

\usepackage{amsxtra}

\usepackage{tensor}

\usepackage[main=british]{babel}

\usepackage[autostyle=false,autopunct=false,english=british]{csquotes}
\setquotestyle{american}
\newcommand*{\defquote}[1]{`#1'}
\newcommand*{\mathquote}[1]{`\,#1\,'}

\usepackage{amsthm}
\makeatletter
\def\@endtheorem{\endtrivlist}
\makeatother

\theoremstyle{remark}

\newtheoremstyle{innote}{\parsep}{\parsep}{\footnotesize}{}{}{}{0pt}{}
\theoremstyle{innote}

\usepackage[shortlabels,inline]{enumitem}
\SetEnumitemKey{para}{itemindent=\parindent,leftmargin=0pt,listparindent=\parindent,parsep=0pt,itemsep=\topsep}
\setlist{itemsep=0pt,topsep=\parsep}
\setlist[enumerate,2]{label=\alph*.}
\setlist[enumerate]{label=\arabic*.,leftmargin=1.5\parindent}
\setlist[itemize]{leftmargin=1.5\parindent}
\setlist[description]{leftmargin=1.5\parindent}
\usepackage[babel,theoremfont,largesc]{newpxtext}

\usepackage[bigdelims,nosymbolsc%,smallerops % probably arXiv doesn't have it
]{newpxmath}
\makeatletter
\def\re@DeclareMathSymbol#1#2#3#4{%
    \let#1=\undefined
    \DeclareMathSymbol{#1}{#2}{#3}{#4}}
\re@DeclareMathSymbol{\bigoplusop}{\mathop}{largesymbols}{"4C}
\re@DeclareMathSymbol{\bigotimesop}{\mathop}{largesymbols}{"4E}
\re@DeclareMathSymbol{\sumop}{\mathop}{largesymbols}{"50}
\re@DeclareMathSymbol{\prodop}{\mathop}{largesymbols}{"51}
\re@DeclareMathSymbol{\bigcupop}{\mathop}{largesymbols}{"53}
\re@DeclareMathSymbol{\bigcapop}{\mathop}{largesymbols}{"54}
\re@DeclareMathSymbol{\bigwedgeop}{\mathop}{largesymbols}{"56}
\re@DeclareMathSymbol{\bigveeop}{\mathop}{largesymbols}{"57}
\re@DeclareMathSymbol{\bigtimesop}{\mathop}{largesymbolsPXA}{"10}
\makeatother
\DeclareFontFamily{U}{egreek}{\skewchar\font'177}%
\DeclareFontShape{U}{egreek}{m}{n}{<-6>s*[1]eurm5 <6-8>s*[1]eurm7 <8->s*[1]eurm10}{}%
\DeclareFontShape{U}{egreek}{m}{it}{<->s*[1]eurmo10}{}%
\DeclareFontShape{U}{egreek}{b}{n}{<-6>s*[1]eurb5 <6-8>s*[1]eurb7 <8->s*[1]eurb10}{}%
\DeclareFontShape{U}{egreek}{b}{it}{<->s*[1]eurbo10}{}%
\DeclareSymbolFont{egreeki}{U}{egreek}{m}{it}%
\SetSymbolFont{egreeki}{bold}{U}{egreek}{b}{it}% from the amsfonts package
\DeclareSymbolFont{egreekr}{U}{egreek}{m}{n}%
\SetSymbolFont{egreekr}{bold}{U}{egreek}{b}{n}% from the amsfonts package
\DeclareFontFamily{U}{egreekx}{\skewchar\font'177}
\DeclareFontShape{U}{egreekx}{m}{n}{%
       <-7.5>s*[0.9]euex7%
    <7.5-8.5>s*[0.9]euex8%
    <8.5-9.5>s*[0.9]euex9%
    <9.5->s*[0.9]euex10%
}{}
\DeclareSymbolFont{egreekx}{U}{egreekx}{m}{n}
\DeclareMathSymbol{\sumop}{\mathop}{egreekx}{"50}
\DeclareMathSymbol{\prodop}{\mathop}{egreekx}{"51}
\DeclareMathSymbol{\coprodop}{\mathop}{egreekx}{"60}
\makeatletter
\def\sum{\DOTSI\sumop\slimits@}
\def\prod{\DOTSI\prodop\slimits@}
\def\coprod{\DOTSI\coprodop\slimits@}
\makeatother
\DeclareMathSymbol{\varpartial}{\mathalpha}{egreeki}{"40}
\DeclareMathSymbol{\partialup}{\mathalpha}{egreekr}{"40}
\DeclareMathSymbol{\alpha}{\mathalpha}{egreeki}{"0B}
\DeclareMathSymbol{\beta}{\mathalpha}{egreeki}{"0C}
\DeclareMathSymbol{\gamma}{\mathalpha}{egreeki}{"0D}
\DeclareMathSymbol{\delta}{\mathalpha}{egreeki}{"0E}
\DeclareMathSymbol{\epsilon}{\mathalpha}{egreeki}{"0F}
\DeclareMathSymbol{\zeta}{\mathalpha}{egreeki}{"10}
\DeclareMathSymbol{\eta}{\mathalpha}{egreeki}{"11}
\DeclareMathSymbol{\theta}{\mathalpha}{egreeki}{"12}
\DeclareMathSymbol{\iota}{\mathalpha}{egreeki}{"13}
\DeclareMathSymbol{\kappa}{\mathalpha}{egreeki}{"14}
\DeclareMathSymbol{\lambda}{\mathalpha}{egreeki}{"15}
\DeclareMathSymbol{\mu}{\mathalpha}{egreeki}{"16}
\DeclareMathSymbol{\nu}{\mathalpha}{egreeki}{"17}
\DeclareMathSymbol{\xi}{\mathalpha}{egreeki}{"18}
\DeclareMathSymbol{\omicron}{\mathalpha}{egreeki}{"6F}
\DeclareMathSymbol{\pi}{\mathalpha}{egreeki}{"19}
\DeclareMathSymbol{\rho}{\mathalpha}{egreeki}{"1A}
\DeclareMathSymbol{\sigma}{\mathalpha}{egreeki}{"1B}
 \DeclareMathSymbol{\tau}{\mathalpha}{egreeki}{"1C}
\DeclareMathSymbol{\upsilon}{\mathalpha}{egreeki}{"1D}
\DeclareMathSymbol{\phi}{\mathalpha}{egreeki}{"1E}
\DeclareMathSymbol{\chi}{\mathalpha}{egreeki}{"1F}
\DeclareMathSymbol{\psi}{\mathalpha}{egreeki}{"20}
\DeclareMathSymbol{\omega}{\mathalpha}{egreeki}{"21}
\DeclareMathSymbol{\varepsilon}{\mathalpha}{egreeki}{"22}
\DeclareMathSymbol{\vartheta}{\mathalpha}{egreeki}{"23}
\DeclareMathSymbol{\varpi}{\mathalpha}{egreeki}{"24}
\let\varrho\rho 
\let\varsigma\sigma
 \let\varkappa\kappa
\DeclareMathSymbol{\varphi}{\mathalpha}{egreeki}{"27}
\DeclareMathSymbol{\varAlpha}{\mathalpha}{egreeki}{"41}
\DeclareMathSymbol{\varBeta}{\mathalpha}{egreeki}{"42}
\DeclareMathSymbol{\varGamma}{\mathalpha}{egreeki}{"00}
\DeclareMathSymbol{\varDelta}{\mathalpha}{egreeki}{"01}
\DeclareMathSymbol{\varEpsilon}{\mathalpha}{egreeki}{"45}
\DeclareMathSymbol{\varZeta}{\mathalpha}{egreeki}{"5A}
\DeclareMathSymbol{\varEta}{\mathalpha}{egreeki}{"48}
\DeclareMathSymbol{\varTheta}{\mathalpha}{egreeki}{"02}
 \DeclareMathSymbol{\varIota}{\mathalpha}{egreeki}{"49}
\DeclareMathSymbol{\varKappa}{\mathalpha}{egreeki}{"4B}
\DeclareMathSymbol{\varLambda}{\mathalpha}{egreeki}{"03}
\DeclareMathSymbol{\varMu}{\mathalpha}{egreeki}{"4D}
\DeclareMathSymbol{\varNu}{\mathalpha}{egreeki}{"4E}
\DeclareMathSymbol{\varXi}{\mathalpha}{egreeki}{"04}
\DeclareMathSymbol{\varOmicron}{\mathalpha}{egreeki}{"4F}
\DeclareMathSymbol{\varPi}{\mathalpha}{egreeki}{"05}
\DeclareMathSymbol{\varRho}{\mathalpha}{egreeki}{"50}
\DeclareMathSymbol{\varSigma}{\mathalpha}{egreeki}{"06}
\DeclareMathSymbol{\varTau}{\mathalpha}{egreeki}{"54}
\DeclareMathSymbol{\varUpsilon}{\mathalpha}{egreeki}{"07}
\DeclareMathSymbol{\varPhi}{\mathalpha}{egreeki}{"08}
\DeclareMathSymbol{\varChi}{\mathalpha}{egreeki}{"58}
\DeclareMathSymbol{\varPsi}{\mathalpha}{egreeki}{"09}
\DeclareMathSymbol{\varOmega}{\mathalpha}{egreeki}{"0A} 
\DeclareMathSymbol{\Alpha}{\mathalpha}{egreekr}{"41}
\DeclareMathSymbol{\Beta}{\mathalpha}{egreekr}{"42}
\DeclareMathSymbol{\Gamma}{\mathalpha}{egreekr}{"00}
\DeclareMathSymbol{\Delta}{\mathalpha}{egreekr}{"01}
\DeclareMathSymbol{\Epsilon}{\mathalpha}{egreekr}{"45}
\DeclareMathSymbol{\Zeta}{\mathalpha}{egreekr}{"5A}
\DeclareMathSymbol{\Eta}{\mathalpha}{egreekr}{"48}
\DeclareMathSymbol{\Theta}{\mathalpha}{egreekr}{"02}
\DeclareMathSymbol{\Iota}{\mathalpha}{egreekr}{"49}
\DeclareMathSymbol{\Kappa}{\mathalpha}{egreekr}{"4B}
\DeclareMathSymbol{\Lambda}{\mathalpha}{egreekr}{"03}
\DeclareMathSymbol{\Mu}{\mathalpha}{egreekr}{"4D}
\DeclareMathSymbol{\Nu}{\mathalpha}{egreekr}{"4E}
\DeclareMathSymbol{\Xi}{\mathalpha}{egreekr}{"04}
\DeclareMathSymbol{\Omicron}{\mathalpha}{egreekr}{"4F}
\DeclareMathSymbol{\Pi}{\mathalpha}{egreekr}{"05}
\DeclareMathSymbol{\Rho}{\mathalpha}{egreekr}{"50}
\DeclareMathSymbol{\Sigma}{\mathalpha}{egreekr}{"06}
\DeclareMathSymbol{\Tau}{\mathalpha}{egreekr}{"54}
\DeclareMathSymbol{\Upsilon}{\mathalpha}{egreekr}{"07}
\DeclareMathSymbol{\Phi}{\mathalpha}{egreekr}{"08}
\DeclareMathSymbol{\Chi}{\mathalpha}{egreekr}{"58}
\DeclareMathSymbol{\Psi}{\mathalpha}{egreekr}{"09}
\DeclareMathSymbol{\Omega}{\mathalpha}{egreekr}{"0A}
\DeclareMathSymbol{\alphaup}{\mathalpha}{egreekr}{"0B}
\DeclareMathSymbol{\betaup}{\mathalpha}{egreekr}{"0C}
\DeclareMathSymbol{\gammaup}{\mathalpha}{egreekr}{"0D}
 \DeclareMathSymbol{\deltaup}{\mathalpha}{egreekr}{"0E}
\DeclareMathSymbol{\epsilonup}{\mathalpha}{egreekr}{"0F}
\DeclareMathSymbol{\zetaup}{\mathalpha}{egreekr}{"10}
\DeclareMathSymbol{\etaup}{\mathalpha}{egreekr}{"11}
\DeclareMathSymbol{\thetaup}{\mathalpha}{egreekr}{"12}
\DeclareMathSymbol{\iotaup}{\mathalpha}{egreekr}{"13}
\DeclareMathSymbol{\kappaup}{\mathalpha}{egreekr}{"14}
\DeclareMathSymbol{\lambdaup}{\mathalpha}{egreekr}{"15}
\DeclareMathSymbol{\muup}{\mathalpha}{egreekr}{"16}
\DeclareMathSymbol{\nuup}{\mathalpha}{egreekr}{"17}
\DeclareMathSymbol{\xiup}{\mathalpha}{egreekr}{"18}
\DeclareMathSymbol{\omicronup}{\mathalpha}{egreekr}{"6F}
  \DeclareMathSymbol{\piup}{\mathalpha}{egreekr}{"19}
\DeclareMathSymbol{\rhoup}{\mathalpha}{egreekr}{"1A}
\DeclareMathSymbol{\sigmaup}{\mathalpha}{egreekr}{"1B}
\DeclareMathSymbol{\tauup}{\mathalpha}{egreekr}{"1C}
\DeclareMathSymbol{\upsilonup}{\mathalpha}{egreekr}{"1D}
\DeclareMathSymbol{\phiup}{\mathalpha}{egreekr}{"1E}
\DeclareMathSymbol{\chiup}{\mathalpha}{egreekr}{"1F}
\DeclareMathSymbol{\psiup}{\mathalpha}{egreekr}{"20}
\DeclareMathSymbol{\omegaup}{\mathalpha}{egreekr}{"21}
\DeclareMathSymbol{\varepsilonup}{\mathalpha}{egreekr}{"22}
\DeclareMathSymbol{\varthetaup}{\mathalpha}{egreekr}{"23}
\DeclareMathSymbol{\varpiup}{\mathalpha}{egreekr}{"24}

\DeclareMathSymbol{\varphiup}{\mathalpha}{egreekr}{"27}
\renewcommand\sfdefault{uop}
\DeclareMathAlphabet{\mathsf}  {T1}{\sfdefault}{m}{sl}
\SetMathAlphabet{\mathsf}{bold}{T1}{\sfdefault}{b}{sl}
\newcommand*{\mathte}[1]{\textbf{\textit{\textsf{#1}}}}
\usepackage[scaled=0.84]{DejaVuSansMono}

\usepackage{mathdots}

\usepackage[usenames]{xcolor}
\definecolor{mypurpleblue}{RGB}{68,119,170}
\definecolor{myblue}{RGB}{102,204,238}
\definecolor{mygreen}{RGB}{34,136,51}
\definecolor{myyellow}{RGB}{204,187,68}
\definecolor{myred}{RGB}{238,102,119}
\definecolor{myredpurple}{RGB}{170,51,119}
\definecolor{mygrey}{RGB}{187,187,187}
\definecolor{lgrey}{RGB}{221,221,221}
\colorlet{shadecolor}{lgrey}

\usepackage{bm}

\usepackage{microtype}

\usepackage[backend=biber,mcite,%subentry,
citestyle=authoryear-comp,bibstyle=pglpm-authoryear,autopunct=false,sorting=ny,sortcites=false,natbib=false,maxcitenames=2,maxbibnames=8,minbibnames=8,giveninits=true,uniquename=false,uniquelist=false,maxalphanames=1,block=space,hyperref=true,defernumbers=false,useprefix=true,sortupper=false,language=british,parentracker=false]{biblatex}
\DeclareSortingTemplate{ny}{\sort{\field{sortname}\field{author}\field{editor}}\sort{\field{year}}}
\DeclareFieldFormat{postnote}{#1}
\DefineBibliographyExtras{british}{\def\finalandcomma{\addcomma}}
\renewcommand*{\finalnamedelim}{\addspace\amp\space}

\DeclareDelimFormat{multicitedelim}{\addsemicolon\addspace\space}
\DeclareDelimFormat{compcitedelim}{\addsemicolon\addspace\space}
\DeclareDelimFormat{postnotedelim}{\addspace}
\ifarxiv\else\fi
\renewcommand{\bibfont}{\footnotesize}
\defbibheading{bibliography}[\bibname]{\section*{#1}\addcontentsline{toc}{section}{#1}%\markboth{#1}{#1}
}
\newcommand*{\citep}{\footcites}
\providecommand{\href}[2]{#2}

\newcommand*{\amp}{\&}

\newcommand*{\subtitleproc}[1]{}

\def\myUrlOrds{\do\0\do\1\do\2\do\3\do\4\do\5\do\6\do\7\do\8\do\9\do\a\do\b\do\c\do\d\do\e\do\f\do\g\do\h\do\i\do\j\do\k\do\l\do\m\do\n\do\o\do\p\do\q\do\r\do\s\do\t\do\u\do\v\do\w\do\x\do\y\do\z\do\A\do\B\do\C\do\D\do\E\do\F\do\G\do\H\do\I\do\J\do\K\do\L\do\M\do\N\do\O\do\P\do\Q\do\R\do\S\do\T\do\U\do\V\do\W\do\X\do\Y\do\Z}%
\makeatletter
\g@addto@macro{\UrlBreaks}{\myUrlOrds}
\makeatother
\newcommand*{\arxiveprint}[1]{%
arXiv \doi{10.48550/arXiv.#1}%
}
\newcommand*{\mparceprint}[1]{%
\href{http://www.ma.utexas.edu/mp_arc-bin/mpa?yn=#1}{mp\_arc:\allowbreak\nolinkurl{#1}}%
}
\newcommand*{\haleprint}[1]{%
\href{https://hal.archives-ouvertes.fr/#1}{\textsc{hal}:\allowbreak\nolinkurl{#1}}%
}
\newcommand*{\philscieprint}[1]{%
\href{http://philsci-archive.pitt.edu/archive/#1}{PhilSci:\allowbreak\nolinkurl{#1}}%
}
\newcommand*{\doi}[1]{%
\href{https://doi.org/#1}{\textsc{doi}:\allowbreak\nolinkurl{#1}}%
}
\newcommand*{\biorxiveprint}[1]{%
bioRxiv \doi{10.1101/#1}%
}
\newcommand*{\osfeprint}[1]{%
Open Science Framework \doi{10.31219/osf.io/#1}%
}
\newcommand*{\osfproj}[1]{%
Open Science Framework \doi{10.17605/osf.io/#1}%
}

\usepackage{graphicx}

\PassOptionsToPackage{hyphens}{url}\usepackage[hypertexnames=false,pdfencoding=unicode,psdextra]{hyperref}

\usepackage[depth=4]{bookmark}
\hypersetup{colorlinks=true,bookmarksnumbered,pdfborder={0 0 0.25},citebordercolor={0.2667 0.4667 0.6667},citecolor=mypurpleblue,linkbordercolor={0.6667 0.2 0.4667},linkcolor=myredpurple,urlbordercolor={0.1333 0.5333 0.2},urlcolor=mygreen,breaklinks=true,pdftitle={\pdftitle},pdfauthor={\pdfauthor}}

\usepackage[british]{datetime2}
\DTMnewdatestyle{mydate}%
{% definitions
}
\DTMsetdatestyle{mydate}

\ifafour\setstocksize{297mm}{210mm}%{*}% A4
\else\setstocksize{210mm}{5.5in}%{*}% 210x139.7
\fi
\settrimmedsize{\stockheight}{\stockwidth}{*}
\setlxvchars[\normalfont] %313.3632pt for a 66-characters line
\setxlvchars[\normalfont]
\ifafour\settypeblocksize{*}{32pc}{1.618} % A4
\else\settypeblocksize{*}{26pc}{1.618}% nearer to a 66-line newpx and preserves GR
\fi
\setulmargins{*}{*}{1}%gives equal margins
\setlrmargins{*}{*}{*}
\setheadfoot{\onelineskip}{2.5\onelineskip}
\setheaderspaces{*}{2\onelineskip}{*}
\setmarginnotes{2ex}{10mm}{0pt}
\checkandfixthelayout[nearest]
\pdfinterwordspaceon%

\newcommand*{\asudedication}[1]{%
{\par\centering\textit{#1}\par}}
\newenvironment{acknowledgements}{\section*{Thanks}\addcontentsline{toc}{section}{Thanks}}{\par}
\makeatletter\renewcommand{\appendix}{\par
  \bigskip{\centering
   \interlinepenalty \@M
   \normalfont
   \printchaptertitle{\sffamily\appendixpagename}\par}
  \setcounter{section}{0}%
  \gdef\@chapapp{\appendixname}%
  \gdef\thesection{\@Alph\c@section}%
  \anappendixtrue}\makeatother
\counterwithout{section}{chapter}
\setsecnumformat{\upshape\csname the#1\endcsname\quad}
\setsecheadstyle{\large\bfseries\sffamily%
\centering}
\setsubsecheadstyle{\bfseries\sffamily%
\raggedright}
\setsubsecindent{0pt}%0ex plus 1ex minus .2ex}
\setparaheadstyle{\bfseries\sffamily%
\raggedright}
\newcommand{\addchap}[1]{\chapter*[#1]{#1}\addcontentsline{toc}{chapter}{#1}}
\newcommand{\addsec}[1]{\section*{#1}\addcontentsline{toc}{section}{#1}}

\copypagestyle{manaart}{plain}
\makeheadrule{manaart}{\headwidth}{0.5\normalrulethickness}
\makeoddhead{manaart}{%
{\footnotesize%\sffamily%
\scshape\headauthor}}{}{{\footnotesize\sffamily%
\headtitle}}
\makeoddfoot{manaart}{}{\thepage}{}

\definecolor{mygray}{gray}{0.333}
\iftypodisclaim%
\ifafour\newcommand\addprintnote{\begin{picture}(0,0)%
\put(245,149){\makebox(0,0){\rotatebox{90}{\tiny\color{mygray}\textsf{This
            document is designed for screen reading and
            two-up printing on A4 or Letter paper}}}}%
\end{picture}}% A4
\else\newcommand\addprintnote{\begin{picture}(0,0)%
\put(176,112){\makebox(0,0){\rotatebox{90}{\tiny\color{mygray}\textsf{This
            document is designed for screen reading and
            two-up printing on A4 or Letter paper}}}}%
\end{picture}}\fi%afourtrue
\makeoddfoot{plain}{}{\makebox[0pt]{\thepage}\addprintnote}{}
\else
\makeoddfoot{plain}{}{\makebox[0pt]{\thepage}}{}
\fi%typodisclaimtrue
\makeoddhead{plain}{\scriptsize\reporthead}{}{}
\pretitle{\begin{center}\LARGE\sffamily%
\bfseries}
\posttitle{\bigskip\end{center}}

\makeatletter\newcommand*{\atf}{\includegraphics[%trim=1pt 1pt 0pt 0pt,
totalheight=\heightof{@}]{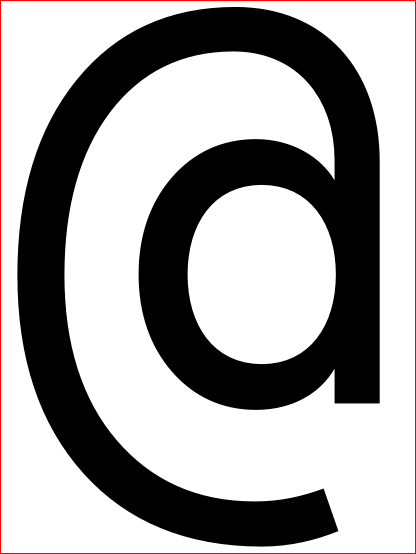}}\makeatother

\providecommand{\epost}[1]{\texttt{\footnotesize\textless#1\textgreater}}
\providecommand{\email}[2]{\href{mailto:#1ZZ@#2 ((remove ZZ))}{#1\protect\atf#2}}

\preauthor{\vspace{-0.5\baselineskip}\begin{center}
\normalsize\sffamily%
\lineskip  0.5em}
\postauthor{\par\end{center}}
\predate{\DTMsetdatestyle{mydate}\begin{center}\footnotesize}
\postdate{\end{center}\vspace{-\medskipamount}}

\setfloatadjustment{figure}{\footnotesize}
\captiondelim{\quad}
\captionnamefont{\footnotesize\sffamily%
}
\captiontitlefont{\footnotesize}
\midsloppy
\paragraphfootnotes%
\footmarkstyle{\textsuperscript{%\color{myred}
\scriptsize\bfseries%\underline{
#1%}
}~}
\title{\propertitle}
\author{%
\hspace*{\stretch{1}}%
\parbox{1\linewidth}%\makebox[0pt][c]%
{\protect\centering P.G.L.  Porta Mana \href{https://orcid.org/0000-0002-6070-0784}{\protect\includegraphics[scale=0.16]{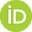}}\\%
\footnotesize Western Norway University of Applied Sciences, Bergen\quad\epost{\email{pgl}{portamana.org}}}%
\hspace*{\stretch{1}}%
}

\date{\firstpublished; updated \updated}

\newcommand*{\de}{\partialup}%partial diff
\newcommand*{\pu}{\piup}%constant pi
\newcommand*{\delt}{\deltaup}%Kronecker, Dirac
\newcommand*{\I}{\mathrm{i}}%imaginary unit
\newcommand*{\e}{\mathrm{e}}%Neper
\newcommand*{\di}{\mathrm{d}}%differential
\DeclareMathOperator{\tr}{tr}%trace
\newcommand*{\incr}{\triangle}%finite increment
\newcommand*{\defd}{\coloneqq}

\DeclarePairedDelimiter\clcl{[}{]}
\DeclarePairedDelimiter\abs{\lvert}{\rvert}
\DeclarePairedDelimiter\set{\{}{\}}
\newcommand*{\p}{\mathrm{P}}%probability
\newcommand*{\sect}{\S}% Sect.~
\newcommand*{\sects}{\S\S}% Sect.~
\newcommand*{\chap}{ch.}%
\newcommand*{\chaps}{chs}%
\newcommand*{\eqn}{eq.}%
\newcommand*{\eqns}{eqs}%
\newcommand*{\ie}{{i.e.}}
\newcommand*{\eg}{{e.g.}}

\newcommand*{\cf}{{cf.}}
\newcommand*{\etal}{{et al.}}

\newcommand*{\tint}{\begingroup\textstyle\int\endgroup\nolimits}
\newcommand*{\T}{^\intercal}%transpose
 \definecolor{notecolour}{RGB}{68,170,153}

\makeatletter
\newcommand*{\q}{}% Check if undefined
\DeclareRobustCommand*{\q}{%
  \mathord{\mathpalette\bigcdot@{}}% changed mathbin to mathord
}
\newcommand*{\bigcdot@scalefactor}{0.7}
\newcommand*{\bigcdot@widthfactor}{1.5}
\newcommand*{\bigcdot@}[2]{%
  \sbox0{$#1\vcenter{}$}% math axis
  \sbox2{$#1\cdot\m@th$}%
  \hbox to \bigcdot@widthfactor\wd2{%
    \hfil
    \raise\ht0\hbox{%
      \scalebox{\bigcdot@scalefactor}{%
        \lower\ht0\hbox{$#1\bullet\m@th$}%
      }%
    }%
    \hfil
  }%
}
\makeatother

\newcommand*{\Un}{\textsf{1}}
\newcommand*{\Le}{\textsf{L}}
\newcommand*{\Ti}{\textsf{T}}
\newcommand*{\Ma}{\textsf{M}}
\newcommand*{\Te}{\Theta}
\newcommand*{\Cu}{\textsf{I}}
\newcommand*{\Ch}{\textsf{Q}}
\newcommand*{\Fl}{\Phi}
\newcommand*{\Ent}{\textsf{s}}
\newcommand*{\En}{\Epsilon}
\newcommand*{\Xx}{\textsf{X}}

\newcommand*{\Aa}{\textsf{A}}

\newcommand*{\Ss}{\textsf{S}}
\newcommand*{\Li}{\textsf{L}}
\newcommand*{\ii}{\cdot}
\newcommand*{\rii}{\ii}
\newcommand*{\yA}{\mathte{A}}
\newcommand*{\yAg}{\mkern2mu\underline{\mkern-2mu\yA\mkern-2mu}\mkern2mu}
\newcommand*{\yB}{\mathte{B}}
\newcommand*{\yBg}{\overline{\yB}}
\newcommand*{\yg}{\mathte{g}}
\newcommand*{\ygc}{\mathte{g}}
\newcommand*{\yT}{\bm{\mathcal{T}}}
\newcommand*{\yTe}{\mathte{T}}
\newcommand*{\yTu}{\bar{\mathte{T}}}
\newcommand*{\yG}{\mathte{G}}
\newcommand*{\yR}{\mathte{Rie}}
\newcommand*{\yRi}{\mathte{Ric}}
\newcommand*{\ysc}{R}
\newcommand*{\yTo}{\bm{\tau}}
\newcommand*{\yom}{\bm{\omega}}
\newcommand*{\yta}{\bm{\xi}}
\newcommand*{\yv}{\bm{v}}
\newcommand*{\yu}{\bm{u}}
\newcommand*{\yw}{\bm{w}}
\renewcommand*{\i}{\indices}
\newcommand*{\dex}[1][i]{\frac{\de}{\de x^{#1}}}
\newcommand*{\dix}[1][i]{\di x^{#1}}
\newcommand*{\nab}{\nabla}
\newcommand*{\yGa}{\varGamma}
\newcommand*{\ds}{\mathit{ds}}
\newcommand*{\inct}{\incr t}
\newcommand*{\yC}{\dot{C}}
\newcommand*{\yU}{\bm{U}}
\newcommand*{\yUu}{\mkern2mu\underline{\mkern-2mu\yU\mkern-2mu}\mkern2mu}
\newcommand*{\id}{\mathbf{id}}%id matrix

\newcommand*{\ygv}{\bm{\gamma}}

\newcommand*{\yk}{\kappa}

\newcommand*{\en}{\epsilon}
\newcommand*{\yq}{q}
\newcommand*{\yp}{p}
\newcommand*{\yt}{\sigma}
\newcommand*{\ytp}{\tau}
\newcommand*{\yl}{\beta}
\newcommand*{\yV}{\bm{V}}
\newcommand*{\dexyz}{\de^3_{xyz}}
\newcommand*{\detzy}{\de^3_{tzy}}
\newcommand*{\detxz}{\de^3_{txz}}
\newcommand*{\detyx}{\de^3_{tyx}}
\newcommand*{\dixyz}{\di^3 xyz}
\newcommand*{\ditzy}{\di^3 tzy}
\newcommand*{\ditxz}{\di^3 txz}
\newcommand*{\dityx}{\di^3 tyx}

\newcommand*{\rul}{{\mkern2mu\rule[-0.1ex]{0.75pt}{1.1ex}\mkern2mu}}
\DeclarePairedDelimiter\mul{\rul}{\rul}%{{\bm{\shortmid}}}
\begin{document}
\captiondelim{\quad}\captionnamefont{\footnotesize}\captiontitlefont{\footnotesize}
\selectlanguage{british}\frenchspacing
\maketitle

%%%%%%%%%%%%%%%%%%%%%%%%%%%%%%%%%%%%%%%%%%%%%%%%%%%%%%%%%%%%%%%%%%%%%%%%%%%%
%%% Abstract
%%%%%%%%%%%%%%%%%%%%%%%%%%%%%%%%%%%%%%%%%%%%%%%%%%%%%%%%%%%%%%%%%%%%%%%%%%%%
\abstractrunin
\abslabeldelim{}
\renewcommand*{\abstractname}{}
\setlength{\absleftindent}{0pt}
\setlength{\absrightindent}{0pt}
\setlength{\abstitleskip}{-\absparindent}
\begin{abstract}\labelsep 0pt%
  \noindent This note provides a short guide to dimensional analysis in
  Lorentzian and general relativity and in differential geometry. It tries
  to revive Dorgelo and Schouten's notion of \defquote{intrinsic} or
  \defquote{absolute} dimension of a tensorial quantity. The intrinsic
  dimension is independent of the dimensions of the coordinates and
  expresses the physical and operational meaning of a tensor. The
  dimensional analysis of several important tensors and tensor operations
  is summarized. In particular it is shown that the components of a tensor
  need not have all the same dimension, and that the Riemann (once
  contravariant and thrice covariant), Ricci (twice covariant), and
  Einstein (twice covariant) curvature tensors are dimensionless. The
  relation between dimension and operational meaning for the metric and
  stress-energy-momentum tensors is discussed; and the possible conventions
  for the dimensions of these two tensors and of Einstein's constant $\yk$,
  including the curious possibility $\yk = 8\pu G$ without $c$ factors, are
  reviewed.
  % \\\noindent\emph{\footnotesize Note: Dear Reader \amp\ Peer,
  %   this manuscript is being peer-reviewed by you. Thank you.}
% \par%\\[\jot]
% \noindent
% {\footnotesize PACS: ***}\qquad%
% {\footnotesize MSC: ***}%
%\qquad{\footnotesize Keywords: ***}
\end{abstract}
\selectlanguage{british}\frenchspacing

%%%%%%%%%%%%%%%%%%%%%%%%%%%%%%%%%%%%%%%%%%%%%%%%%%%%%%%%%%%%%%%%%%%%%%%%%%%%
%%% Epigraph
%%%%%%%%%%%%%%%%%%%%%%%%%%%%%%%%%%%%%%%%%%%%%%%%%%%%%%%%%%%%%%%%%%%%%%%%%%%%
 \asudedication{\small per la piccola Emma}
% \vspace{\bigskipamount}
% \setlength{\epigraphwidth}{.7\columnwidth}
% %\epigraphposition{flushright}
% \epigraphtextposition{flushright}
% %\epigraphsourceposition{flushright}
% \epigraphfontsize{\footnotesize}
% \setlength{\epigraphrule}{0pt}
% %\setlength{\beforeepigraphskip}{0pt}
% %\setlength{\afterepigraphskip}{0pt}
% \epigraph{\emph{text}}{source}

%%%%%%%%%%%%%%%%%%%%%%%%%%%%%%%%%%%%%%%%%%%%%%%%%%%%%%%%%%%%%%%%%%%%%%%%%%%%
%%% BEGINNING OF MAIN TEXT
%%%%%%%%%%%%%%%%%%%%%%%%%%%%%%%%%%%%%%%%%%%%%%%%%%%%%%%%%%%%%%%%%%%%%%%%%%%%

\section{Introduction}
\label{sec:intro}

From the point of view of dimensional analysis, do all components of a
tensor need to have the same dimension? What happens to these components if
we choose coordinates that don't all have dimensions of length or time? And
if the components of a tensor have different dimensions, then does it make
sense to speak of \enquote{the dimension of the tensor}? What are the
dimensions of the metric and of the curvature tensors? What is the
dimension of the constant in the Einstein equations?

A sense of insecurity gets hold of many students (and possibly of some
researchers) in relativity, when they have to discuss and answer this kind
of questions. % the dimensions
% of tensors and of tensor components, the effect of tensor operators on
% dimensions, and the dimensions of constants in differential-geometric and
% field equations.
This is evident in many question \amp\ answer websites and wiki pages,
where several incorrect or unfounded statements about dimensional analysis
in relativity are in circulation.\footnote{The answers to the questions above are:
 \begin{enumerate*}[label=\arabic*.]
\item No, they don't. \item See \eqn~\eqref{eq:dim_component_generic}
  below. \item Yes. \item They are dimensionless. \item
  $\Ma^{-1}\Le^{-1}\Ti^{2}$, or $\Ma^{-1}\Le$, or
  $\Ma^{-1}\Le^{3}\Ti^{-2}$, depending on the dimensions you assign to the
  metric and stress-energy-momentum tensors, see
  \sects~\ref{sec:metric}--\ref{sec:einstein_eq}.
\end{enumerate*}}
% (The answers to the
% questions above are:
% \begin{enumerate*}[label=\arabic*.]
% \item No, they don't. \item See \eqn~\eqref{eq:dim_component_generic}
%   below. \item Yes. \item They are dimensionless. \item
%   $\Ma^{-1}\Le^{-1}\Ti^{2}$, or $\Ma^{-1}\Le$, or
%   $\Ma^{-1}\Le^{3}\Ti^{-2}$, depending on the dimensions you assign to the
%   metric and stress-energy-momentum tensors, see
%   \sects~\ref{sec:metric}--\ref{sec:einstein_eq}.
% \end{enumerate*}
% )
% For example, the notion that the components
% of a tensor or the coordinate functions should all have the same dimension.
% As we'll see shortly, this notion is false.

Several factors contribute to these misconceptions and insecurity. Modern
texts in Lorentzian and general relativity commonly use geometrized units.
They say that, for finding the dimension of some constant in a tensorial
equation, it's sufficient to compare the dimensions of the terms in the
equation. But the application of this procedure is sometimes not so
immediate, because some tensors don't have universally agreed dimensions --
prime example the metric tensor. Many texts use four coordinates with
dimension of length, and base their dimensional analyses on that specific
choice \citep[\eg][p.~71 \eqn~(37.1)]{tolman1934_t1949}[p.~80
\eqn~(32.15)]{landauetal1939_t1996},
% [p.~332 \eqn~(10.15)]{adleretal1965_r1975}[\sect~5.7]{penfieldetal1967}
multiplying timelike tensorial components by appropriate powers of $c$ (there are exceptions of course \citep[\eg][]{dedonder1925,dedonder1926}[\sect~V.55]{fock1955_t1964}[\sect~4.1]{mcvittie1956_r1965}[\sect~VII.1]{fokker1960_t1965}[\sect~F.III.280]{truesdelletal1960}[\sect~X]{kitano2013}). Such common practices can therefore give students the impression that coordinates ought to always be lengths, and that all components of a tensor ought to have the same dimension. Yet, students cannot find such rules explicitly stated anywhere. We'll see shortly that no such rules in fact exist, nor are they necessary.

\medskip

Dimensional analysis is thus not very self-evident in relativity and in
differential geometry. The present note wants to provide a short but
exhaustive guide to it.
% the application of the usual rules of dimensional analysis to
% differential geometry. The usual rules are that if two quantities are
% summed they must have the same dimension; that the dimension of a product
% is the product of the dimensions; and so on.
Some important dimensional-analysis questions in general relativity are
also consistently settled here; for example the dimension of the Riemann
curvature tensor, or the effect of the covariant or Lie derivatives on
dimensions.

The application of dimensional analysis in relativity is most
straightforward and self-evident if we rely on the coordinate-free or
intrinsic approach to differential geometry, briefly recalled below, and if
we adopt the perhaps overlooked notion of \emph{intrinsic dimension} of a
tensor. %The present note tries to revive this notion.
The intrinsic dimension of a tensor was
% pivots on the perhaps forgotten notion of \emph{intrinsic
%   dimension} of a tensor. The present note tries to revive this notion,
introduced under the name \defquote{absolute dimension} by Schouten and Dorgelo \citep{dorgeloetal1946}[\chap~VI]{schouten1951_r1989} and used in Truesdell \amp\ Toupin \citep[Appendix II]{truesdelletal1960}, Post \citep{post1982c}, and recently in Hehl \amp\ Obukhov \citep[\sect~B.1]{hehletal2003}{hehletal2004b}. As its name implies, this dimension is independent of the choice and dimensions of coordinate functions. It is distinct from the dimensions of the tensor's \emph{components}, which instead depend on the dimensions of the coordinates. The intrinsic dimension of a tensor is determined by the latter's physical and operational \citep{bridgman1927_r1958} meaning. It is therefore a natural notion for dimensional analysis in relativity.

\medskip

Here is a synopsis of the rest of this note. The intrinsic approach to
differential geometry is outlined, with references, in the next section,
together with some notation necessary to our discussion.
Section~\ref{sec:2d_example} gives a simple example of dimensional analysis
for a two-dimensional spacetime. This example might be enough for most
readers to grasp the basic way of reasoning; such readers can work out the
rest for themselves whenever they need and don't need to read the rest of
this note. Sections~\ref{sec:coords}--\ref{sec:curves} offer a more
systematic discussion and a synopsis of dimensional analysis for the main
tensorial operations. The notion of intrinsic dimension is explained in
\sect~\ref{sec:tensors}. The intrinsic dimensions of various curvature
tensors, of the metric tensor, and of the stress-energy-momentum tensor are
discussed in \sects~\ref{sec:connection}--\ref{sec:stressenergy}. In
particular, the contravariant and thrice covariant Riemann, twice covariant
Ricci, and twice covariant Einstein curvature tensors are found to have
intrinsic dimension $\Un$, that is, to be dimensionless. The operational
motivation of several standard choices for the dimensions of the metric and
stress-energy-momentum tensors are also discussed. The possible dimensions
of the constant in the Einstein equations are finally derived in
\sect~\ref{sec:einstein_eq}.

This note obviously assumes familiarity with basic tensor calculus and
related notions, for example of co- and contra-variance, tensor product,
contraction. Some passages assume familiarity with the exterior calculus of
differential forms. The general ideas, however, should be understandable
even without such familiarity.

% Section~\ref{sec:tensors}--\ref{sec:curves}
% offer a more general and systematic discussion and a synopsis of the main
% tensorial operations. % It
% is shown, in particular, that \emph{the components of a tensor need not
% have the same dimensions}.

% The final \sect~\ref{sec:summary} gives a summary of the main results, with
% some additional comments.

% \medskip

% This note assumes familiarity with basic tensor calculus and related
% notions, for example of co- and contra-variance, tensor product,
% contraction. Some passages assume familiarity with the exterior calculus of
% differential forms. Above all, familiarity with the intrinsic presentation
% of differential geometry mentioned above is assumed; but the general ideas
% should be understandable even without such familiarity.

Finally, quoting Truesdell \amp\ Toupin \citep[Appendix \sect~7
footnote~4]{truesdelletal1960}, \enquote{dimensional analysis remains a
  controversial and somewhat obscure subject. We do not attempt a complete
  presentation here}. References about recent developments in this subject
are given in the summary of \sect~\ref{sec:summary}.

% \citep{aldersley1977} also confuses component- and absolute dimensions

\section{Intrinsic view of differential-geometric objects:\\ brief
  reminder and notation}
\label{sec:remined}

From the intrinsic point of view, a tensor is defined by its geometric
properties. For example, a vector field $\yv$ % \equiv \yv(\dotv)
is an object that operates on functions defined on the (spacetime)
manifold, yielding new functions, with the properties
$\yv(af+bg)=a\yv(f)+b\yv(g)$ and $\yv(fg)=\yv(f)g+f\yv(g)$ for all
functions $f$, $g$ and reals $a$, $b$. A covector field (also called
1-form) $\yom$ is an object that operates on vector fields, yielding
functions, with the property $\yom(f\yu+g\yv)=f\yom(\yu)+g\yom(\yv)$ for
all vector fields $\yu$, $\yv$ and functions $f$, $g$. The sum of vector or
covector fields and their products by functions are defined in an obvious
way. Tensors are constructed from these objects; see also the end of this
section for a slightly different point of view.

A system of coordinates $(x^{i})$ is just a set of linearly independent
functions. This set gives rise to a set of vectors fields
$\bigl(\dex\bigr)$ and to a set of covector fields $(\dix)$ by the obvious
requirements that $\dex(x^{j})=\delt\i{_{i}^{j}}$ and
$\dix\bigl(\dex[j]\bigr)=\delt\i{^{i}_{j}}$. These two sets can be used as bases
to express all other vectors and covectors as linear combinations. A
vector field $\yv$ can thus be written as
\begin{equation}
  \label{eq:vector_intrinsic}
  \yv \equiv \sum_{i}v^{i}\dex \equiv v^{i}\dex \ ,
\end{equation}
where the functions $v^{i}\defd \dix(\yv)$ are its components with respect
to the basis $\bigl(\dex\bigr)$. Analogously for a covector field.

\medskip

For the presentation of the intrinsic view I recommend the excellent texts
by Choquet-Bruhat \etal\ \cite*{choquetbruhatetal1977_r1996}; Boothby
\cite*{boothby1975_r2003}; Abraham \etal\ \cite*{abrahametal1983_r1988};
Burke \cite*{burke1985_r1987}; Bossavit \cite*{bossavit1991}; Gratus \cite*{gratus2017} for insightful pictorial illustrations. More on
the general-relativity side, Misner \etal\
\cite*[\chap~9]{misneretal1970_r1973}; Gourgoulhon
\cite*[\chap~2]{gourgoulhon2007_r2012}; Penrose \amp\ Rindler
\cite*{penroseetal1984_r2003}. \textcolor{white}{If you find this
  you can claim a postcard from me.}

\medskip

For the notation in dimensional analysis I use \textsc{iso}
conventions:\citep[\sect~5]{iso2009} $\dim(\yA)$ is the dimension of the
quantity $\yA$, and among the base quantities are mass $\Ma$, length $\Le$,
time $\Ti$, temperature $\Te$, electric current $\Cu$, and the
dimensionless $\Un$. Note that I don't discuss units -- it doesn't matter
here whether the unit for length is the metre or the furlong, for example.

Throughout this note $c$ denotes the speed of light, with
$\dim(c) = \Le\Ti^{-1}$. Its numerical quantity value $\set{c}$ depends on
the chosen units of length and time.

The number, ordering, and symmetries of a tensor's covariant and
contravariant \enquote{slots} \citep[\sect~3.2]{misneretal1970_r1973} will
be important in our discussion. The traditional coordinate-free notation
\mathquote{$\yA$} omits this information. We thus need a coordinate-free
notation that makes it explicit when needed. Penrose \amp\ Rindler
\citep[\sect~2.2]{penroseetal1984_r2003} propose an abstract-index notation
where \mathquote{$A\i{_{i}^{jk}}$}, for example, denotes a tensor covariant
in its first slot and contravariant in its second and third slots. Every
index in this notation is \enquote{a \emph{label} whose sole purpose is to
  keep track of the type of tensor under
  discussion}~\citep[p.~75]{penroseetal1984_r2003}. So this notation
doesn't stand for a \emph{component} of the tensor. For the latter, Penrose
\amp\ Rindler use \textbf{bold} indices instead:
\mathquote{$A\i{_{\pmb{i}}^{\pmb{j}\pmb{k}}}$}. But in our discussion the
difference between a tensor and its set of components is crucial, and
Penrose \amp\ Rindler's abstract-index notation unfortunately lends itself
to conceptual and typographic misunderstanding.

I shall therefore use a notation such as $\yA\i{_{\q}^{\q\q}}$ to indicate
that $\yA$ is covariant in its first slot and contravariant in its second
and third slots. Its components would thus be $\bigl(A\i{_{i}^{jk}}\bigr)$.
For brevity I'll call this a \defquote{co-contra-contra-variant} tensor,
with an obvious naming generalization for other tensor
types. % In spirit this notation is similar to the
% abstract-index one of Penrose \amp\ Rindler
% \citep[\sect~2.2]{penroseetal1984_r2003}; but I avoid using letters for the
% indices, so that the notational difference between a tensor and its
% components, which is very important for our discussion, doesn't disappear.
A set of completely antisymmetric slots will be put within bars: thus the
notation $\yA\i{^{\q}_{\mul{\q\q}}}$ means that $\yA$ is completely
antisymmetric in its last two covariant slots. Finally, in accord with
convenient modern terminology, completely antisymmetric contravariant
tensors of order $k$ will be called \defquote{$k$-vectors}; and completely
antisymmetric covariant tensors of order $k$, \defquote{$k$-covectors}. The
terms \defquote{multi-vector} and \defquote{multi-covector} are used when
$k$ isn't specified.

The only weak points of the notation just explained are the operations of
transposition and contraction, which the index notation depicts so well
instead. Considering that transposition is a generalization of matrix
transposition, and contraction a generalization of trace, I'll use the
following notation:
\begin{itemize}[wide]%,label=\textemdash]
\item $\yA^{\intercal_{\alpha\beta}}$ is the transposition
  (swapping) of the $\alpha$th and $\beta$th slots. Its coordinate-free
  definition is
  \begin{equation}
    \label{eq:def_transp}
    (\yA^{\intercal_{\alpha\beta}})(\dotso,
    \overset{\mathclap{\scriptstyle\text{$\alpha$th slot}}}{\bm{\zeta}}
    ,\dotso,
    \underset{\mathclap{\scriptstyle\text{$\beta$th slot}}}{\bm{\eta}}
    ,\dotso)
    \defd
    \yA(\dotso,
    \overset{\mathclap{\scriptstyle\text{$\alpha$th slot}}}{\bm{\eta}}
    ,\dotso,
    \underset{\mathclap{\scriptstyle\text{$\beta$th slot}}}{\bm{\zeta}}
    ,\dotso)
  \end{equation}
  for all $\bm{\zeta},\bm{\eta}$ of appropriate variance type.
  
\item $\tr_{\alpha\beta}\yA$ is the contraction of the $\alpha$th and
  $\beta$th slots, which must have opposite variance types; note that we may
  have $\beta < \alpha$. Its coordinate-free definition is
  \begin{equation}
    \label{eq:def_contr}
    \begin{gathered}
      (\tr_{\alpha\beta}\yA)(\dotso
        ,\dotso
        ,\dotso)
        \defd
        \sum_{i} \yA(\dotso,
        \overset{\mathclap{\scriptstyle\text{$\alpha$th slot}}}{\bm{u}_{i}}
        ,\dotso,
        \underset{\mathclap{\scriptstyle\text{$\beta$th slot}}}{\bm{\omega}^{i}}
        ,\dotso)
      \\
      \parbox{0.9\columnwidth}{\small for any arbitrary complete and
        linearly independent sets \;$\set{\bm{u}_{i}}$,
        $\set{\bm{\omega}^{j}}$\\
        such that \;$\bm{\omega}^{j}(\bm{u}_{i}) = \delt\i{^{j}_{i}}$~.}
        % representation\; $\tsum_{i}
        % \bm{u}_{i}\otimes\bm{\omega}^{i} \equiv \id$\; of\\ the identity
        % operator (tensor) on the tangent space, \;$\id\colon
        % \bm{v} \mapsto \bm{v}$ 
    \end{gathered}
  \end{equation}
\end{itemize}
In index notation the two operations above are the familiar
\begin{equation*}
  A\i{_{\dotsm\ }^{%
      \overset{\mathclap{\scriptstyle\text{$\alpha$th slot}}}{i}
    }_{\dotsm\ %
      \underset{\mathclap{\scriptstyle\text{$\beta$th slot}}}{j}
      \ \dotsm}}
  \mapsto
  A\i{_{\dotsm\ %
      \underset{\mathclap{\scriptstyle\text{$\alpha$th slot}}}{j}
      \ \dotsm\ }^{%
      \overset{\mathclap{\scriptstyle\text{$\beta$th slot}}}{i}
    }_{\ \dotsm}} 
  \qquad\text{and}\qquad
  A\i{_{\dotsm\ }^{%
      \overset{\mathclap{\scriptstyle\text{$\alpha$th slot}}}{i}
    }_{\dotsm\ %
      \underset{\mathclap{\scriptstyle\text{$\beta$th slot}}}{i}
      \ \dotsm}} \ .
\end{equation*}

For the sake of notation economy I'll denote the contraction of adjacent
slots of two tensors by simple juxtaposition. For example, if
$\yA \equiv \yA\i{_{\q}^{\q}}$, $\yB \equiv \yB\i{_{\q\q}}$, and $\yv$ is a
vector, then
\begin{equation}
  \label{eq:simple_contraction}
  \yA\yB \defd \tr_{23}(\yA\i{_{\q}^{\q}}\otimes\yB\i{_{\q\q}})\ ,
  \qquad
  \yB\yv \defd \tr_{23}(\yB\i{_{\q\q}} \otimes \yv\i{^{\q}}) \ .
\end{equation}
This notation makes sense considering tensors as linear operators.

Contraction and transposition will be discussed only sparsely, so I hope
you won't find the notation above too uncomfortable.

% $\yA\i{^{\q \mathrm{c}}_{\q \mathrm{c}\q}}$

\bigskip

%\paragraph{Multi-vectors, multi-covectors, orientations }

It is possible to build the tensor-product architecture not on vectors and covectors, but on multi-vectors and multi-covectors, with their straight and twisted (also called \defquote{even} and \defquote{odd}, %\citep{derham1955_t1984}
or \defquote{polar} and \defquote{axial}% \citep{truesdelletal1960}
) orientations. This elegant and powerful geometric point of view leads to deeper physical insights and is gaining popularity in the literature. For its presentation I recommend the texts of Bossavit \cite*[especially \chap~3]{bossavit1991}; Burke \cite*{burke1983,burke1985_r1987,burke1980b,burke1995}; de~Rham \cite*[\chap~2]{derham1955_t1984}; Schouten \cite*{schouten1924_r1954}; Cartan \cite*[\chap~I]{cartan1928_t1983}; Deschamps \cite*{deschamps1970,deschamps1981}; Lindell \cite*{lindell2004}; Gratus \cite*{gratus2017}.

In the notation above, the bars identify $k$-vectors and $k$-covectors for
$k>1$. Thus $\yA\i{^{\q}_{\mul{\q\q}}}$ indicates that $\yA$ belongs to the
tensor product of 1-vectors and 2-covectors; it's also called a
vector-valued 2-covector. To avoid burdening the notation I won't add
symbols denoting straight or twisted orientation, but I'll explicitly state
in the text when any object has a twisted orientation.

% $\yA\i{_{\q}^{\ori{\q\q}}}$

% $\yA\i{_{\q}^{\q\mulb{\q\q}}}$

% $\yA\i{_{\q}^{\q\mul{\q\q}}}$

% $\yA\i{_{\q\q\mulb{\q\q}}}$

% $\yA\i{_{\q\q\mul{\q\q}}}$

% $\yA\i{_{\q\q\mulc{\q\q}}}$

\section{An introductory two-dimensional example}
\label{sec:2d_example}

Let me first present a simple example of dimensional analysis for a
two-dimensional spacetime. I provide very little explanation, letting the
analysis speak for itself. The next sections will give a longer discussion
of the general point of view, of the assumptions, and of cases with more
elaborate geometric objects.

In a region of a two-dimensional spacetime we use coordinates $(x,y)$.
These coordinates allow us to uniquely label every event in the region
(otherwise they wouldn't be coordinates). Let us say that coordinate $x$
has dimension of temperature, and $y$ of specific entropy:
\begin{equation}
  \label{eq:example_coords2d}
  \dim(x)=\Te \ ,\quad
  \dim(y)= \Ent \defd \Le^{2}\Ti^{-2}\Te^{-1} \ .
\end{equation}
This choice could be possible for several reasons. For example, the region
could be occupied by a heat-conducting material; in a specific spacetime
foliation, its temperature increases along each 1-dimensional spacelike
slice, and its entropy density is uniform on each slice but increases from
slice to slice. \citep[For general-relativistic thermomechanics see
\eg][]{eckart1940c,maugin1974b,maugin1978b,maugin1978c,maugin1978d,maugin1978e,muschiketal2014}
Owing to this kind of monotonic behaviour for these quantities, if we are
given a pair of temperature \amp\ specific-entropy values we can identify a
unique event associated to them in this spacetime region. They can thus be
used as a coordinate system. The point here is that coordinates can have
any dimensions, because of physical reasons. In atmospheric and ocean
dynamics, for example, pressure or mass density are sometimes used as
coordinates for depth
\citep[\chap~6]{griffies2004}[\sect~2.6.2]{vallis2006}.

From these coordinates we construct two covector fields $(\di x, \di y)$,
and two vector fields $\bigl(\frac{\de}{\de x}, \frac{\de}{\de y}\bigr)$
that serve as bases for the spaces of tangent covectors, vectors, and
tensors. Their dimensions are
\begin{equation}
  \label{eq:example_basisvects2d}
  \begin{aligned}
    \dim(\di x)&=\Te &\quad \dim(\di y)&= \Ent \ ,
    \\
    \dim\Bigl(\frac{\de}{\de x}\Bigr)&=\Te^{-1} &\quad \dim\Bigl(\frac{\de}{\de y}\Bigr)&= \Ent^{-1} \ .
  \end{aligned}
\end{equation}

Consider a contra-co-variant tensor field $\yA \equiv \yA\i{^{\q}_{\q}}$ in
this region. Using the basis fields above it can be written as
\begin{equation}
  \label{eq:expansion_tensor_2d}
  \yA = 
  A\i{^{x}_{x}}\,\frac{\de}{\de x} \otimes\di x + 
  A\i{^{x}_{y}}\,\frac{\de}{\de x} \otimes\di y + 
  A\i{^{y}_{x}}\,\frac{\de}{\de y} \otimes\di x + 
  A\i{^{y}_{y}}\,\frac{\de}{\de y} \otimes\di y \ ,
\end{equation}
where $A\i{^{x}_{x}} \defd \yA\bigl(\di x, \frac{\de}{\de x}\bigr)$,
$A\i{^{x}_{y}} \defd \yA\bigl(\di x, \frac{\de}{\de y}\bigr)$, and so on,
are the components of the tensor in the coordinate system $(x, y)$.

By the rules of dimensional analysis, the two sides of the expansion above,
and in fact each summand on the right side, must have the same dimension.
Denoting $\Aa \defd \dim(\yA)$, we thus find the four equations
\begin{equation*}
  \label{eq:four_dims_2d_pre}
    \Aa = \dim(A\i{^{x}_{x}}) =
 \dim(A\i{^{x}_{y}})\,\Te^{-1}\,\Ent 
 =
 \dim(A\i{^{y}_{x}})\,\Te\,\Ent^{-1}
 =
 \dim(A\i{^{y}_{y}}) \ ,
\end{equation*}
or
\begin{equation}
  \label{eq:four_dims_2d}
  \begin{aligned}
    \dim(A\i{^{x}_{x}}) &= \Aa &\qquad 
 \dim(A\i{^{x}_{y}}) &=\Aa\,\Te\,\Ent^{-1} \equiv
\Aa\,\Le^{-2}\Ti^{2}\Te^{2}
    \\
    \dim(A\i{^{y}_{x}}) &= \Aa\,\Te^{-1}\,\Ent &\qquad
 \dim(A\i{^{y}_{y}}) &= \Aa \ .
  \end{aligned}
\end{equation}

The intrinsic dimension of the tensor $\yA$ is $\Aa$. The
expansion~\eqref{eq:expansion_tensor_2d} shows that this dimension is
independent of the coordinate system, by construction -- such expansion
could be done in any other coordinate system, and the left side would be
the same. The effect of coordinate transformations is examined more in
detail in \sect~\ref{sec:tensors}. The intrinsic dimension $\Aa$ is
determined by the physical and operational meaning of the tensor $\yA$; see
\sects~\ref{sec:metric}, \ref{sec:stressenergy} for concrete examples.
Together with the dimensions of the coordinates it determines the
dimensions of the components, \eqns~\eqref{eq:four_dims_2d}, which need not
be all equal.

This simple example should have disclosed the main points of dimensional
analysis on manifolds, which will now be discussed in more generality. In
the derivation above we silently adopted a couple of natural conventions:
for example, that the tensor product behaves similarly to multiplication
with regard to dimensions. Such conventions are briefly discussed in
\sect~\ref{sec:summary}.

\section{Coordinates}
\label{sec:coords}

From a physical point of view a coordinate is just a function that
associates values of some physical quantity with the events in a region
(the domain of the coordinate chart) of spacetime. Together with the other
coordinates such function allows us to uniquely identify every event
within that region. Any physical quantity will do: the distance from
something, the time elapsed since something, an angle, an energy density,
the strength of a magnetic flux, a temperature, and so on. A coordinate can
thus have any dimension: length $\Le$, time $\Ti$, angle $\Un$, temperature
$\Te$, magnetic flux $\Fl \defd \Ma\Le^{2}\Ti^{-2}\Cu^{-1}$, and so on.

The functional relation between two sets of coordinates must of course be
dimensionally consistent. For example, if $\dim(x^{0})=\Ti$ and
$\dim(x^{1})=\Le$, and we introduce a new coordinate $y(x^{0},x^{1})$
with dimension $\Fl$, additive in the previous two, then we must have
$y = a x^{0} + b x^{1}$ with $\dim(a) = \Fl\Ti^{-1}$ and
$\dim(b) = \Fl\Le^{-1}$.

%The dimensions of the coordinates don't matter, as we'll now see.

\section{Tensors: intrinsic dimension and components' dimensions}
\label{sec:tensors}

Consider a system of coordinates $(x^i)$ with dimensions $(\Xx_i)$, and the
ensuing sets of covector fields $(\dix)$ and of vector fields
$\bigl(\dex\bigr)$, bases for the cotangent and tangent spaces. Their
tensor products are bases for the tangent spaces of higher tensor types.

The differential $\dix$ traditionally has the same dimension as $x^{i}$:
$\dim(\dix) = \Xx_{i}$, and the vector $\dex$ traditionally has the
inverse dimension: $\dim\bigl(\dex\bigr) = {\Xx_{i}}^{-1}$.
% We'll see later that these conventions are self-consistent.

For our discussion let's take a concrete example: a contra-co-variant tensor
field $\yA \equiv \yA\i{^{\q}_{\q}}\,$. The discussion generalizes to tensors
of other types in an obvious way.

The tensor $\yA$ can be expanded in terms of the basis vectors and
covectors, as in \sect~\ref{sec:remined} and in the example of
\sect~\ref{sec:2d_example}:
\begin{equation}
  \label{eq:expansion_tensor}
  \yA = A\i{^{i}_{j}}\, \dex\otimes\dix[j]
  \equiv A\i{^{0}_{0}}\,\dex[0]\otimes\dix[0] + 
  A\i{^{0}_{1}}\,\dex[0]\otimes\dix[1] + \dotsb {}\ .
\end{equation}
Each function
\begin{equation}
  A\i{^{i}_{j}} \defd  \yA\Bigl(\dix, \dex[j]\Bigr)
  \equiv \tr_{12}\tr_{34}\Bigl(\yA\i{^{\q}_{\q}} \otimes \dix \otimes
  \dex[j]\Bigr)
  \equiv \dix\,\yA\,\dex[j]
  \label{eq:components_def}
\end{equation}
is a component of the tensor in this coordinate system.

\medskip

To make dimensional sense, all terms in the sum~\eqref{eq:expansion_tensor}
must have the same dimension. This is possible only if the generic
component $A\i{^{i}_{j}}$ has dimension
\begin{equation}
  \label{eq:dim_component}
  \dim\bigl(A\i{^{i}_{j}}\bigr) = \Aa\ {\Xx_{i}}\,{\Xx_{j}}^{-1}\ ,
\end{equation}
where $\Aa$ is common to all components. In fact, the
${\Xx_{i}}\,{\Xx_{j}}^{-1}$ term cancels the ${\Xx_{i}}^{-1}\,{\Xx_{j}}$
term coming from $\dex\otimes\dix[j]$ in the
sum~\eqref{eq:expansion_tensor}, and each summand therefore has dimension
$\Aa$.

The generalization of the formula above to tensors of other types is obvious:
\begin{equation}
  \label{eq:dim_component_generic}
\dim\bigl(A\i{^{ij\dotso}_{kl\dotso}}\bigr) = \Aa\ {\Xx_{i}}\,{\Xx_{j}}\dotsm\,
  {\Xx_{k}}^{-1}\,{\Xx_{l}}^{-1}\dotsm \ ,
\end{equation}
where the ordering of the indices doesn't matter.
% For example, if  we're using coordinates with dimensions
% \begin{equation}
%   \label{eq:example_coords}
%   \dim(x^{0})=\Te\ ,\quad
%   \dim(x^{1})=\Le\ ,\quad
%   \dim(x^{2})=\Le\ ,\quad
%   \dim(x^{3})=\Ma\Le^{-1}\Ti^{-2}\ ,
% \end{equation}
% then the components of $\yA$ have dimensions
% \begin{equation}
%   \label{eq:example_components}
%   \Bigl(\dim\bigl(A\i{^{i}_{j}}\bigr)\Bigr) =
%  \Aa\times \begin{pmatrix}
%    \Un & \Le^{-1}\Te & \Le^{-1}\Te & \Ma^{-1}\Le\Ti^{2}\Te
%    \\
%    \Le\Te^{-1} & \Un & \Un & \Ma^{-1}\Le^{2}\Ti^{2}
%    \\
%    \Le\Te^{-1} & \Un & \Un & \Ma^{-1}\Le^{2}\Ti^{2}
%    \\
%    \Ma\Le^{-1}\Ti^{-2}\Te^{-1} & \Ma\Le^{-2}\Ti^{-2} & \Ma\Le^{-2}\Ti^{-2} & \Un
%   \end{pmatrix}\ .
% \end{equation}
Clearly the components can have different dimensions \citep[\cf\ the
discussion in][\sect~IV.5 p.~179]{synge1960b}. What matters is that the
sum~\eqref{eq:expansion_tensor} be dimensionally consistent.

\medskip

The dimension $\Aa$, which is also the dimension of the
sum~\eqref{eq:expansion_tensor}, I'll call the \emph{intrinsic dimension}
of the tensor $\yA$, and we write
\begin{equation}
  \label{eq:abs_dim}
  \dim(\yA) = \Aa\ .
\end{equation}
This dimension is independent of any coordinate system. It reflects the
physical or operational \citep{bridgman1927_r1958}[see
also][\sect~A.2]{synge1960}[\sects~A.3--4]{truesdelletal1960} meaning of
the tensor. We shall see an example of such an operational analysis in
\sects~\ref{sec:metric}, \ref{sec:stressenergy} for the metric and
stress-energy-momentum tensors.

The notion of intrinsic dimension was introduced by Dorgelo and
Schouten~\citep{dorgeloetal1946}[\chap~VI]{schouten1951_r1989} under the
name \defquote{absolute dimension}. I find the adjective
\defquote{intrinsic} more congruous to modern terminology and less prone to
suggest spurious connections with absolute values.
In the following I'll drop the adjective \defquote{intrinsic} when it is
clear from the context.

\medskip

Different coordinate systems lead to different dimensions of the
\emph{components} of a tensor $\yA$, but the intrinsic dimension of the
tensor remains the same. Formula~\eqref{eq:dim_component_generic} for the
dimensions of the components is consistent under changes of coordinates.
For example, in new coordinates $({\bar{x}}^{k})$ with dimensions
$({\bar{\Xx}}_{k})$, the new components of $\yA$ are
\begin{equation}
  \label{eq:coords_change}
  \bar{A}\i{^{k}_{l}} = A\i{^{i}_{j}}
  \,\frac{\de \bar{x}^{k}}{\de x^{i}}
  \,\frac{\de {x}^{j}}{\de \bar{x}^{l}}
\end{equation}
and a quick check shows that
$\dim(\bar{A}\i{^{k}_{l}}) = \Aa\,\bar{\Xx}_{k}\,{\bar{\Xx}_{l}}^{-1}$, consistently
with the general formula~\eqref{eq:dim_component_generic}. % (You may have noticed
% that the partial-derivative symbol $\de$ is not in boldface in the last
% expression; that's on purpose but I won't dwell on the reason here, because
% it doesn't affect our discussion.)

\medskip

If in \eqn~\eqref{eq:coords_change}, relating intrinsic and component
dimensions, all coordinates have equal dimensions, $\Xx_{i} =\Xx$ for all
$i$, then all components also have equal dimensions. So \emph{if we use a
  system of coordinates having equal dimensions, the components of any
  tensor must also have equal dimensions}. This justifies common practice
in the literature.

Choosing coordinates of different dimensions, however, has several
advantages. First, it allows us to use dimensional analysis as a heuristic
tool to determine the variance type of a tensor; we'll see an example in
\sect~\ref{sec:stressenergy}. Second, it can lead to components with
familiar dimensions. For example, if we use a timelike coordinate of
dimension $\Ti$ and spacelike coordinates of dimension $\Le$, then the
components of the (co-contra-variant) stress-energy-momentum tensor have
the familiar dimensions of energy density, surface energy-flux density,
momentum density, and pressure, with no $c$ factors involved; see again
\sect~\ref{sec:stressenergy}.

% the timelike and spacelike components of the charge-current
% form have dimensions $\Cu\Le^{-3}\Ti$ and $\Cu\Le^{-2}$ -- the familiar
% charge density and surface current density.

\section{Tensor operations}
\label{sec:tensor_ops}

By the reasoning of the previous section, which simply applies standard
dimensional considerations to the basis
expansion~\eqref{eq:expansion_tensor}, it's easy to find the resulting
intrinsic dimension of various operations and operators on tensors and
tensor fields.

Here is a summary of the dimensional rules for the main
differential-geometric operations and operators, except for the covariant
derivative, the metric, and related tensors, discussed more in depth in
\sects~\ref{sec:connection}--\ref{sec:metric} below. Some of these rules
are actually definitions or conventions, as briefly discussed in their
description. The others can be proved; I only give proofs for some of them,
leaving the other proofs as an exercise. For reference, in brackets I give
the section of Choquet-Bruhat \etal\ \cite*{choquetbruhatetal1977_r1996}
where these operations are defined.

\begin{itemize}[wide=0pt]
\item The \emph{tensor product} [III.B.5] multiplies dimensions:
  \begin{equation}
  \dim(\yA\otimes\yB) = \dim(\yA)\dim(\yB)\ .\label{eq:tensor_mult}
\end{equation}

This is actually a definition or convention. We tacitly used this rule
already in the example of \sect~\ref{sec:2d_example} and in
\sect~\ref{sec:tensors} for the coordinate
expansion~\eqref{eq:expansion_tensor}. It is a natural definition, because
for tensors of order 0 (functions) the tensor product is just the ordinary
product, and the dimension of a product is the product of the dimensions.
This definition doesn't lead to inconsistencies.

% Consider for example the tensor product of $\yA\i{^{\q}_{\q}}$ and
% $\yB\i{_{\q\q}^{\q}}$. It introducing coordinates $(x^{i})$ and their
% related basis vectors and covectors, it can be written as the sum
% \begin{equation}
%   \label{eq:tensor_prod_example}
%   \yA \otimes \yB =
%   A\i{^{i}_{j}}\,B\i{_{kl}^{m}}\ 
%   \dex[i]\otimes\dix[j]\otimes\dix[k]\otimes\dix[l]\otimes\dex[m]\ .
% \end{equation}
% Equating the dimensions of the left and right sides, and considering that
% \begin{equation}\label{eq:dim_A_and_B}
%   \dim(A\i{^{i}_{j}})= \dim(\yA)\,\Xx_{i}\,{\Xx_{j}}^{-1}\ ,
%   \quad
%   \dim(B\i{_{kl}^{m}})= \dim(\yB)\,{\Xx_{k}}^{-1}\,{\Xx_{l}}^{-1}\,\Xx_{m}\ ,
% \end{equation}
% owing to the generalization of the formula for the
% components~\eqref{eq:dim_component_generic}, we find that all $\Xx$ terms cancel
% out, leaving the result~\eqref{eq:tensor_mult}.

% \smallskip

% from which it follows that
% \begin{equation}
%   \label{eq:dim_comp_tensor_prod}
%   \dim(A\i{^{i}_{j}}\,B\i{_{kl}^{m}}) =
%   \Aa\,\Bb\ 
%   \Xx_{i}\,{\Xx_{j}}^{-1}\,{\Xx_{k}}^{-1}\,{\Xx_{l}}^{-1}\,\Xx_{m}
% \end{equation}
% with $\Aa = \dim(\yA)$ and $\Bb = \dim(\yB)$. Comparing with the
% generalization of the formula for the components~\eqref{eq:dim_component_generic},
% we see that the intrinsic dimension of $\yA\otimes\yB$ is therefore
% $\Aa\Bb \equiv \dim(\yA)\,\dim(\yB)$.

\item The \emph{contraction} [III.B.5] or trace of the $\alpha$th and $\beta$th
  slots of a tensor has the same dimension as the tensor:
  \begin{equation}
    \dim(\tr_{\alpha\beta}\yA) = \dim(\yA)\ .
    \label{eq:tensor_contr}
  \end{equation}
  Note that the formula above only holds \emph{without raising or lowering
    indices}; see \sect~\ref{sec:g_inv_vol} for those operations.

  This operation can be traced back to the duality of vectors and covectors
  mentioned in \sect~\ref{sec:remined}: a covector field $\yom$ operates on
  a vector field $\yv$ to yield a function $f=\yom(\yv)$. Also in this case
  we have that $\dim(f)=\dim(\yom)\dim(\yv)$ by definition or convention,
  and the rule~\eqref{eq:tensor_contr} follows from this convention. Also
  in this case this convention seems very natural, owing to the linearity
  properties of the trace, and doesn't lead to inconsistencies.

\item The \emph{transposition} \citep[called \enquote{building an isomer}
  by][\sect~I.3 p.~13]{schouten1924_r1954}[\sect~II.4
  p.~20]{schouten1951_r1989} of the $\alpha$th and $\beta$th slots of a
  tensor has the same dimension as the tensor:
  \begin{equation}
    \dim(\yA^{\intercal_{\alpha\beta}}) = \dim(\yA)\ .
    \label{eq:tensor_transp}
  \end{equation}

\item The \emph{Lie bracket} [III.B.3] of two vectors has the product of their dimensions:
  \begin{equation}
    \dim(\clcl{\yu,\yv}) =\dim(\yu)\dim(\yv)\ .
    \label{eq:lie_bracket}
\end{equation}

In fact, in coordinates $(x^{i})$ the bracket can be expressed as
\begin{equation}
  \label{eq:bracket_coords}
  \clcl{\yu,\yv} =
  \biggl( u\i{^{j}}\frac{\de v\i{^{i}}}{\de x\i{^{j}}}
  - v\i{^{j}}\frac{\de u\i{^{i}}}{\de x\i{^{j}}} \biggr)\ \dex\ ,
\end{equation}
and equating the dimensions of the left and right sides, considering that
\begin{equation}
  \label{eq:dim_u_and_v}
  \dim(u\i{^{i}}) = \dim(\yu)\,\Xx_{i}\ ,\quad
  \dim(v\i{^{i}}) = \dim(\yv)\,\Xx_{i}\ ,
\end{equation}
we find again that all $\Xx$ terms cancel out, leaving the
result~\eqref{eq:lie_bracket}.

\smallskip

\item The \emph{pull-back} [III.A.2], \emph{tangent map} [III.B.1], and
  \emph{push-forward} of a map $F$ between manifolds don't change the
  dimensions of the tensors they map. The reason, evident from their
  definitions, is that they all rest on the pull-back of a function:
  $F^{*}(f) \defd f\circ F$, which, being a composition, has the same
  dimension as the function.
%   \begin{equation}
%   \dim(\yA\otimes\yB) = \dim(\yA)\dim(\yB)\ .\label{eq:tensor_mult}
% \end{equation}

\smallskip

\item The \emph{Lie derivative} [III.C.2] of a tensor with respect to a
  vector field has the product of the dimensions of the two:
  \begin{equation}
    \dim(\Li_{\yv}\yA) =\dim(\yv)\dim(\yA) \ .
    \label{eq:lie_der}
\end{equation}
\end{itemize}

\bigskip

Regarding operations and operators on differential forms:

\begin{itemize}[wide=0pt]
\item The \emph{exterior product} [IV.A.1] of two differential forms
  multiplies their dimensions:
  \begin{equation}
  \dim(\yom\land\yta) = \dim(\yom)\dim(\yta) \ .\label{eq:ext_prod}
\end{equation}
  
\item The \emph{interior product} [IV.A.4] (also called \defquote{inner},
  \defquote{dual}, or \defquote{dot} product) of a vector and a covector
  multiplies their dimensions:
  \begin{equation}
    \dim(\yv\ii\yom) =\dim(\yv)\dim(\yom) \ .
    \label{eq:inter_prod}
\end{equation}
This equation also holds for the generalized interior product
\citep{deschamps1970}[Appendices]{deschamps1981}{lindell2004}[\sect~F.I.267]{truesdelletal1960}[Box
4.1, item 4]{misneretal1970_r1973}[see also][]{portamana2019e} of a
multi-vector $\yv$ and a multi-covector $\yom$. The interior product is
also often denoted \mathquote{$\mathrm{i}_{\yv}\yom$} or
\mathquote{$\yv\mathbin{\!\rfloor\!}\yom$}.

\item The \emph{exterior derivative} [IV.A.2] of a form has the same
  dimension of the form:
  \begin{equation}
    \dim(\di\yom) =\dim(\yom) \ .
    \label{eq:ext_deriv}
  \end{equation}
  This can be proven using the identity
  $\di(\yv\ii\yom)+\yv\ii(\di\yom) = \Li_{\yv}\yom$ or similar identities
  \citep[\chap~9 p.~180 Theorem~9.78]{curtisetal1985}[\sect~6.4
  Theorem~6.4.8]{abrahametal1983_r1988} together with
  \eqns~\eqref{eq:lie_der} and~\eqref{eq:inter_prod}.

\item The \emph{integral} [IV.B.1] of a form over a submanifold (or more
  generally a chain) $M$ has the same dimension as the form:
  \begin{equation}
    \dim\bigl(\tint_{M}\yom\bigr) =\dim(\yom) \ .
    \label{eq:integration}
  \end{equation}
\end{itemize}
The reason is that the integral of a form over a submanifold or chain
ultimately rests on the standard definition of integration on the real line
\citep[\eg][\sects~IV.B.1--2]{choquetbruhatetal1977_r1996}[\sect~5 p.~21,
\sect~6
p.~24]{derham1955_t1984}[\sect~7.1]{abrahametal1983_r1988}[\sect~VI.2]{boothby1975_r2003},
which satisfies the dimensional rule above. In fact, the integral is
invariant with respect to reparameterizations of the chain; it depends only
on its image (some texts~\citep[\eg][\sect~10.4]{martin1991_r2004}[\sect~7.3]{fecko2006} even define chains as
equivalence classes determined by their image).

\medskip

% The resultant intrinsic dimensions of other operators, for example the
% determinant \citep[\sect~6.2]{abrahametal1983_r1988}, can be obtained by
% similar reasoning.

All rules above extend in obvious ways to tensor densities, and apply
regardless whether the objects have straight or twisted orientations.

% extend in obvious ways to inner-oriented forms
% \citep[\chap~II]{schouten1951_r1989} (also called \defquote{odd}
% \citep[\chap~II]{derham1955_t1984} or \defquote{twisted}
% \citep{burke1983,burke1995}[\chap~3]{bossavit1991} forms) and to .

\section{Curves and integral curves}
\label{sec:curves}

Consider a curve into spacetime $C\colon s \mapsto P(s)$, with the
parameter $s$ having some dimension $\dim(s)=\Ss$.

If we consider the events of the spacetime manifold as dimensionless
quantities, then the dimension of the tangent or velocity vector $\yC$
to the curve is
\begin{equation}
  \label{eq:dim_velocity}
  \dim(\yC) = \Ss^{-1} 
\end{equation}
owing to the definition
\citep[\sect~III.B.1]{choquetbruhatetal1977_r1996}[\sect~IV.(1.9)]{boothby1975_r2003}
\begin{equation}
  \label{eq:def_tangent_curve}
\yC \defd \frac{\de (x^{i} \circ C)}{\de s}\ \dex \ .
\end{equation}
% or considering that $\yC$ can be interpreted as the push-forward of
% $\partial_s$, that is, $C_*(\partial_s)$.

Note an important consequence of this fact. Given a vector field $\yv$ we
say that $C$ is an integral curve for it if
\begin{equation}
  \yv = \yC
  \label{eq:integral_curve}
\end{equation}
at all events $C(s)$ in the image of the curve (or
$\yv_{C(s)} = \yC_{C(s)}$ in standard differential-geometric notation
\citep[\sect~III.B.1]{choquetbruhatetal1977_r1996}). From the point of view
of dimensional analysis this definition is only valid if $\yv$ has
dimension $\Ss^{-1}$. If $\yv$ and $s^{-1}$ have different dimensions -- a
case which could happen for physical reasons -- the
condition~\eqref{eq:integral_curve} must be modified into $\yv = k\yC$,
where $k$ is a dimensionful constant. This is equivalent to considering an
affine and dimensional reparameterization of $C$.

Worldlines and their 4-velocities are discussed in \sect~\ref{sec:g_4velocity}.

\section{Connection, covariant derivative, curvature tensors}
\label{sec:connection}

Consider an arbitrary connection
\citep[\sect~V.B]{choquetbruhatetal1977_r1996} with covariant derivative
$\nab$. For the moment we don't assume the presence of any metric
structure.

The covariant derivative of the product $f\yv$ of a function and a vector
satisfies \citep[\sect~V.B.1]{choquetbruhatetal1977_r1996}
\begin{equation}
  \label{eq:basic_property_covder}
  \nab(f\yv) = \di f \otimes \yv + f\nab\yv\ .
\end{equation}
The first summand, from formulae~\eqref{eq:ext_deriv}
and~\eqref{eq:tensor_mult}, has dimension $\dim(f)\dim(\yv)$; for
dimensional consistency this must also be the dimension of the second
summand. Thus
\begin{equation}
  \label{eq:dim_cov_der_vect}
  \dim(\nab\yv) = \dim(\yv)\ .
\end{equation}
It follows that the \emph{directional} covariant derivative $\nab_{\yu}$
has dimension
\begin{equation}
  \label{eq:dim_dircov_der_vect}
  \dim(\nab_{\yu}\yv) = \dim(\yu)\dim(\yv)\ ,
\end{equation}
and by its derivation properties \citep[\sect~V.B.1
p.~303]{choquetbruhatetal1977_r1996} we see that
formula~\eqref{eq:dim_cov_der_vect} extends from vectors to 
tensors of arbitrary type.

\medskip

In the coordinate system $(x^{i})$ the action of the covariant derivative
is carried by the \emph{connection coefficients} or Christoffel symbols
$\bigl(\yGa\i{^{i}_{jk}}\bigr)$ defined by
\begin{equation}
  \label{eq:christoffel}
  \nab\dex[k] = \yGa\i{^{i}_{jk}}\  \dix[j]\otimes\dex[i]\ .
\end{equation}
From this equation and \eqns~\eqref{eq:tensor_mult},
\eqref{eq:dim_cov_der_vect} it follows that an individual coefficient has
dimension
\begin{equation}
  \label{eq:dim_christoffel}
  \dim\bigl(\yGa\i{^{i}_{jk}}\bigr) = \Xx_{i}\, {\Xx_{j}}^{-1}\,{\Xx_{k}}^{-1}\ .
\end{equation}

\medskip

The \emph{torsion} $\yTo \equiv \yTo\i{^{\q}_{\mul{\q\q}}}$, \emph{Riemann
  curvature} $\yR \equiv \yR\i{^{\q}_{\q\mul{\q\q}}}$, and \emph{Ricci
  curvature} $\yRi \equiv \yRi\i{_{\q\q}}$ tensors are defined by
\citep[\sect~V.B.1]{choquetbruhatetal1977_r1996}
\begin{gather}
  \label{eq:torsion_eq}
\yTo(\yu,\yv) \defd \nab_{\yu}\yv - \nab_{\yv}\yu - \clcl{\yu,\yv}\ ,
\\
\yR(\yw;\yu,\yv) \defd
\nab_{\yu}\nab_{\yv}\yw - \nab_{\yv}\nab_{\yu}\yw
- \nab_{\clcl{\yu,\yv}}\yw\ ,
  \label{eq:riemann_eq}  
\\
\yRi\i{_{\q\q}} \defd \tr_{13} \yR\i{^{\q}_{\q\mul{\q\q}}} \ .
  \label{eq:ricci_eq}  
\end{gather}
From these definitions and the results of \sect~\ref{sec:tensor_ops} we
find the dimensional requirements
\begin{gather}
  \label{eq:torsion_eq_dim}
\dim\bigl(\yTo\i{^{\q}_{\mul{\q\q}}}\bigr)\dim(\yu)\dim(\yv) = \dim(\yu) \dim(\yv)\ ,
\\
\dim\bigl(\yR\i{^{\q}_{\q\mul{\q\q}}}\bigr)\dim(\yw)\dim(\yu)\dim(\yv) =
\dim(\yw)\dim(\yu)\dim(\yv)\ ,
\label{eq:riemann_eq_dim}
\\
\dim(\yRi\i{_{\q\q}}) = \dim\bigl(\yR\i{^{\q}_{\q\mul{\q\q}}}\bigr)\ ,
  \label{eq:ricci_eq_dim}  
\end{gather}
which imply that the \emph{torsion, Riemann curvature, and Ricci curvature
  tensors are dimensionless}:
\begin{equation}
  \label{eq:dim_torsion_riemann_ricci}
  \dim\bigl(\yTo\i{^{\q}_{\mul{\q\q}}}\bigr) =
  % \label{eq:dim_riem}
   \dim\bigl(\yR\i{^{\q}_{\q\mul{\q\q}}}\bigr) =
  % \label{eq:dim_ricci}
   \dim(\yRi\i{_{\q\q}}) =\Un\ .
\end{equation}
This result is sensible because the notion of local parallelism, which
these tensors express, doesn't involve any notion of distance or angle
\citep[\cf][]{portamana2011_r2019}. The exact contra- and co-variant
primitive type of these tensors is very important in the equations above.
If a metric tensor is also introduced and used to raise or lower any
indices of these tensors, the resulting tensors will have different
dimensions; see \sect~\ref{sec:g_inv_vol}.

The result~\eqref{eq:dim_torsion_riemann_ricci} appears in Post \citep[\sect~8]{post1982c}. Post also states that the \emph{intrinsic} dimension of the connection coefficients is unity, which could seemingly be at variance with \eqn~\eqref{eq:dim_christoffel}; but I have not managed to understand which intrinsic geometric object he associates with the connection coefficients. If that object is the covariant derivative $\nab$, then his statement agrees with \eqn~\eqref{eq:dim_cov_der_vect}.

Misner \etal\ \citep[p.~35%, 407
]{misneretal1970_r1973} say that \enquote{curvature}, by which they seem to
mean the Riemann tensor, has dimension $\Le^{-2}$. This statement is
seemingly at variance with the dimensionless
results~\eqref{eq:dim_torsion_riemann_ricci}. But I believe that Misner
\etal\ refer to the \emph{components} of the Riemann tensor in
\emph{specific coordinates} of dimension $\Le$. In such specific
coordinates every \emph{component} $\mathit{Rie}\i{^{i}_{jkl}}$ has
dimension $\Le^{-2}$, according to the general
formula~\eqref{eq:dim_component_generic}, if and only if the intrinsic
dimension of $\yR$ is unity, $\dim(\yR)=\Un$. So I believe that Misner
\etal's statement actually agrees with the
results~\eqref{eq:dim_torsion_riemann_ricci}. This possible
misunderstanding shows the importance of distinguishing between the
intrinsic dimension, which doesn't depend on any specific coordinate
choice, and component dimensions, which do.

The formulae above are also valid if a metric is defined and the connection
is compatible with it, see \sect~\ref{sec:g_einst} below.

\section{Metric and related tensors and operations}
\label{sec:metric}

\subsection{Intrinsic dimensions: two choices}
\label{sec:g_intr_dim}

Let us now consider a metric tensor $\yg \equiv \yg\i{_{\q\q}}$. What is
its intrinsic dimension $\dim(\yg)$? The literature offers two choices;
both can be motivated by the operational meaning of the metric.

Consider a (timelike) worldline $s \mapsto C(s)$, $s\in \clcl{a,b}$,
between events $C(a)$ and $C(b)$. The metric tells us the \emph{proper
  time} $\inct$ elapsed for an observer having that worldline, according to
the formula
%\begin{subequations}\label{eq:dimg_t}\tag{(\ref{eq:dimg_t})}
\begin{equation}%\addtocounter{equation}{1}\tag{\theequation a}
  \label{eq:proper_time}
\inct =  \int_a^b
\sqrt{\abs*{\yg[\yC(s),\yC(s)]}} \  \di s\ .
\end{equation}
From the results of \sect~\ref{sec:tensor_ops} this formula implies that
%\begin{equation}
%  \label{eq:dim_dt}
 $\Ti \equiv \dim(\inct) = \sqrt{\dim(\yg\i{_{\q\q}})}$,
%\end{equation}
 independently of the dimension of the parameter $s$, and therefore
\begin{equation}%\addtocounter{equation}{1}\tag{\theequation a}
  \label{eq:dim_g}
  \dim(\yg\i{_{\q\q}}) = \Ti^{2}\ .
\end{equation}
%\end{subequations}

Most authors \citep[\eg][\sect~V.62 \eqn~(62.02)]{fock1955_t1964}[\chap~11
\eqn~(11.21)]{curtisetal1985}[\sect~5.3
\eqn~(5.6)]{rindler1969_r1986}[\chap~6 \eqn~(6.24)]{hartle2003}, however,
prefer to include a dimensional factor $1/c$ in the
definition~\eqref{eq:proper_time}:
\begin{equation}%\addtocounter{equation}{-1}\tag{\theequation b}
  \label{eq:proper_time_c}
  \inct = \int_a^b
  \frac{1}{c} \sqrt{\abs*{\ygc[\yC(s),\yC(s)]}} \ \di s\ ,
\end{equation}
thus obtaining
\begin{equation}%\addtocounter{equation}{1}\tag{\theequation b}
  \label{eq:dim_g_c}
  \dim(\ygc\i{_{\q\q}}) = \Le^{2}\ .
\end{equation}
%To avoid confusion I'm using $\ygc$ for the metric with such dimension.

The choice~\eqref{eq:dim_g_c} is also supported by the traditional
expression for the \enquote{line element $\ds^{2}$} as it appears in many
works:
\begin{equation}
  \label{eq:line_elem}
  \ds^{2} = -c^2\mathit{dt}^2 + \mathit{dx}^2 +\mathit{dy}^2 + \mathit{dz}^2
  \ ,
\end{equation}
sometimes with opposite sign. If the coordinates $(t,x,y,z)$ have the
dimensions suggested by their symbols, this formula has dimension
$\Le^{2}$, so that if we interpret \enquote{$\ds^{2}$} as $\ygc\i{_{\q\q}}$
we find $\dim(\ygc\i{_{\q\q}}) = \Le^{2}$. The line-element expression
above often has an ambiguous differential-geometric meaning, however,
because it may also represent the metric applied to some \emph{unspecified}
vector, that is, $\ds^{2} = \yg(\yv,\yv)$ with $\yv$ left unspecified
\citep[\cf][Box~3.2~D p.~77]{misneretal1970_r1973}. In this case we have
\begin{equation*}
  \label{eq:line_elem_dim_ambiguous}
  \Le^{2} = \dim(\yg)\,\dim(\yv)^{2}
\end{equation*}
and the dimension of $\yg$ is ambiguous or undefined, because the vector
$\yv$ could have any dimension.% -- possibly dimensionless if $\yv$
% has dimension of length, but we'll see in \sect~\ref{sec:einstein_eq} that
% a dimensionless $\yg$ isn't quite compatible with the Einstein equation.

\medskip

The standard choices for $\dim(\yg)$ are thus $\Le^{2}$ or $\Ti^{2}$; the
corresponding metric tensors differ by a factor $c^{2}$.

The choice $\dim(\yg) \defd \Ti^{2}$, used for example by McVittie, Synge,
Kilmister
\citep[\sect~4.1]{mcvittie1956_r1965}[\sect~IV.5]{synge1960b}[\chap~II
p.~25]{kilmister1973}, has some advantages for the definition of the
co-variant 4-velocity, discussed in \sect~\ref{sec:g_4velocity}. It could
be motivated on operational grounds for reasons discussed by Synge and
Bressan \citep[\sects~III.2--4]{synge1960b}[\sects~15, 18]{bressan1978}.
Synge gives a vivid summary: \citep[\sect~III.3 pp.~108--109]{synge1960b}
\begin{quote}\footnotesize
  We are now launched on the task of giving physical meaning to the
  Riemannian geometry \textelp{}. It is indeed a Riemannian
  \emph{chronometry} rather than \emph{geometry}, and the word
  \emph{geometry}, with its dangerous suggestion that we should go about
  measuring \emph{lengths} with \emph{yardsticks}, might well be abandoned
  altogether in the present connection
\end{quote}
In fact, to measure the proper time $\inct$ defined above we only need to
ensure that a clock has the worldline $C$, and then take the difference
between the clock's final and initial times. %  On the other hand, consider
% the case when the curve $C$ is \emph{spacelike}. Its proper length is
% still defined by the integral~\eqref{eq:proper_time} apart from a
% dimensional constant.
If $C$ is spacelike instead, the measurement of its proper \emph{length},
still defined by the integral~\eqref{eq:proper_time} apart from a
dimensional constant, is more involved. It requires dividing the curve into
very short pieces, and having specially-chosen observers (with 4-velocities
orthogonal to the pieces) measure each piece. To measure each short piece,
each observer uses radar distance, sending a light signal which bounces
back at the end of the piece, and timing how long it takes to come back
\citep[\chap~2]{frankel1979}[\sect~84]{landauetal1939_t1996}. Even if rigid
rods are used, their calibration still relies on a measurement of time --
this is also reflected in the current definition of the standard metre
\citep[p.~98]{bipm1983}[p.~25]{giacomo1984}. Thus the measurement of length
seems to ultimately rely on the measurement of time. It could be objected,
however, that the laws of light propagation depend on the metric tensor,
which connects the Faraday and Maxwell tensors
\citep[\chap~F.III]{truesdelletal1960}[\chap~II.4]{misneretal1970_r1973}{puntigametal1997,hehletal2001,hehletal2004},
so this reasoning could be circular.

The other choice, $\dim(\yg) \defd \Le^{2}$, is by far the most common. It
has the merit that the projection of the metric onto a spacelike
hypersurface also has dimension $\Le^{2}$, which is sensible from a
Newtonian point of view. Such projections are at the heart of $3+1$
formulations of general relativity \citep(I thank I.~Bengtsson for this
remark){gourgoulhon2007_r2012,alcubierre2008}[\chap~21]{misneretal1970_r1973}{wilsonetal2003_r2007,smarretal1978,york1979,smarretal1980}
and also of covariant formulations of Newtonian mechanics
\citep[\sects~B.II.152--154, D.II.203--205, D.V.238,
F.IV.285--289]{truesdelletal1960}[\sect~2.4]{marsdenetal1983b_r1994}.

Post \citep[\sects~5, 8]{post1982c} offers some arguments for a dimensionless metric tensor: $\dim(\yg) \defd \Un$. In particular he states: \citep[\sects~5 p.~183]{post1982c}
\begin{quote}\footnotesize
  Since the relative [\ie, component] dimensions are the primary sources of information relating to measurement, a situation now obtains where dimensions determine index positions of physical tensors, which in turn means physical dimensions determine transformation characteristics. It is the first step for making the principle of covariance unique.
\end{quote}
I disagree with this conclusion. It is based on an implicit choice of coordinates of dimensions $(\Ti,\Le,\Le,\Le)$. With such choice and a dimensionless metric, then raising or lowering an index of a tensor does indeed lead to \emph{components} having different dimensions. But the \emph{intrinsic} dimension of the tensor would be unaffected, as shown in the next section; and even the component dimensions are unaffected if we choose dimensionless coordinates. A dimensionful metric, instead, always leads to a different intrinsic dimension of the tensor with a raised or lowered index, see \eqn~\eqref{eq:raising_lowering}. This is more desirable in coordinate-free physics and geometry.

\medskip

In the following we shall consider the choices $\dim(\yg) \defd \Le^{2}$ and $\dim(\yg) \defd \Ti^{2}$, showing how they affect several dimensional results. % Another merit is that the norm of a 4-velocity -- and therefore
% also its spacelike projection, the ordinary 3-velocity -- has indeed the
% dimension of a velocity, $\Le\Ti^{-1}$.

\subsection{Inverse metric, index raising and lowering, proper volume element}
\label{sec:g_inv_vol}

The metric $\yg$ can be considered as an operator mapping vectors $\yv$ to
covectors $\yom$, which we can compactly write as $\yom=\yg\yv$ as
discussed in \sect~\ref{sec:remined}. The \emph{inverse metric tensor}
$\yg^{-1} \equiv \yg\i{^{-1}^{\q\q}}$ is then defined by
\begin{equation}
  \label{eq:inverse_g}
  \yg\,\yg^{-1} = \id\i{_{\q}^{\q}} \ ,
  \qquad
  \yg^{-1}\,\yg = \id\i{^{\q}_{\q}} \ ,
  % \qquad
  % \yg\yg^{-1} = \id\i{_{\q}^{\q}} = (\id\i{^{\q}_{\q}})^{\intercal}\ ,
\end{equation}
where $\id\i{_{\q}^{\q}}\colon \yom \mapsto \yom$ is the dimensionless
identity operator (also a tensor) on the cotangent space, and
$\id\i{^{\q}_{\q}}$ on the tangent space. Hence
\begin{equation}
  \label{eq:dim_ig}
  \dim(\yg^{-1}) = \dim(\yg)^{-1} \ .
  % \begin{cases}
  %   \Ti^{-2} \\
  %   \Le^{-2} \\
  %   \Un 
  % \end{cases}
  % \text{if }\dim(\yg)\defd
  % \begin{cases}
  %    \Ti^2 \\
  %   \Le^2 \\
  %    \Un 
  % \end{cases}
\end{equation}

\medskip

The operation of \emph{raising or lowering an index} of a tensor represents
a contraction of the tensor product of that tensor with the metric or the
metric inverse, for example
$\yAg\i{_{\q\q}} \defd \yg\yA \equiv \tr_{23}(\yg\i{_{\q\q}} \otimes
\yA\i{^{\q}_{\q}})$ from the tensor $\yA\i{^{\q}_{\q}}$, and similarly for
tensors of other types. Therefore every lowering of a tensor's index
multiplies its dimension by $\dim(\yg)$, and every rising divides it by
$\dim(\yg)$:
\begin{equation}
  \label{eq:raising_lowering}
  \begin{split}
  \dim(\yAg\i{_{\dotso}_{\q}_{\dotso}}) &=
  \dim(\yA\i{_{\dotso}^{\q}_{\dotso}})\,\dim(\yg)
\\
  \dim(\yBg\i{_{\dotso}^{\q}_{\dotso}}) &=
  \dim(\yB\i{_{\dotso}_{\q}_{\dotso}})\,\dim(\yg)^{-1} \ .
\end{split}
\end{equation}

\medskip

% in spacetime is an outer-oriented 4-form $\ygv$, equivalent to a completely
% antisymmetric tensor $\ygv\i{_{\q\q\q\q}}$, such that
% $\ygv(\ye_{0},\ye_{1},\ye_{2},\ye_{3})=1$ for every quadruple of
% positively-oriented orthonormal vector fields $(\ye_{k})$ (orthonormal
% obviously means $\abs{\yg(\ye_{k},\ye_{l})}=\delt_{kl}$). The volume element
% is determined by the metric and by an inner orientation of spacetime, which
% is independent of the metric. It has only one non-zero component, given by
% the square root of the determinant of the %(positively ordered)
% components $(g\i{_{ij}})$ of the metric:

The volume element in a four-dimensional spacetime is a twisted 4-form
uniquely determined by the metric tensor
\citep[\sect~V.24]{derham1955_t1984}[\sect~V.A.4]{choquetbruhatetal1977_r1996}[\sect~6.2]{abrahametal1983_r1988}.
Its only non-zero component is equal to the square root of the determinant
of the %(positively ordered)
components $(g\i{_{ij}})$ of the metric:
\begin{equation*}
  \label{eq:undim_volume_elem}
  \sqrt{\smash[b]{\abs{\det(g\i{_{ij}})}}}\ \dix[0]\land\dix[1]\land\dix[2]\land\dix[3] \ .
\end{equation*}
Here
$\dix[0]\land\dix[1]\land\dix[2]\land\dix[3]$ %the sequence $(x^{0},x^{1}, x^{2}, x^{3})$ must define a positive orientation.
actually has a twisted orientation \citep[\chap~6 p.~60,
\chap~9]{frankel1979} (the coordinate transformation of the non-zero
component includes the sign of the Jacobian), which in this case means that
it has no screw-sense orientation at all, only an abstract \mathquote{$+$}
orientation. For this reason a globally non-vanishing volume element can be
defined on orientable and non-orientable manifolds alike. From the results
of \sect~\ref{sec:tensor_ops} it can be shown that the 4-form above has
intrinsic dimension $\dim(\yg)^{2}$ (in an $n$-dimensional spacetime it has
dimension $\dim(\yg)^{n/2}$). It's convenient to multiply it by a power of
$c$ and to define the \emph{proper volume element}
$\ygv \equiv \ygv\i{_{\mul{\q\q\q\q}}}$ as follows:
\begin{equation}
  \label{eq:volume_elem}
  \ygv \defd
      \begin{cases}
        \frac{1}{c}\
        \sqrt{\smash[b]{\abs{\det(g\i{_{ij}})}}}\ \dix[0]\land\dix[1]\land\dix[2]\land\dix[3] &\quad  \text{if }\dim(\yg)\defd    \Le^2 \\[\jot]
        c^{3}\
        \sqrt{\smash[b]{\abs{\det(g\i{_{ij}})}}}\ \dix[0]\land\dix[1]\land\dix[2]\land\dix[3] &\quad  \text{if }\dim(\yg)\defd
     \Ti^2
  \end{cases}.
\end{equation}
As a consequence we have
\begin{equation}
  \label{eq:dim_volume_elem}
  \dim\bigl(\ygv\i{_{\mul{\q\q\q\q}}}\bigr) = \Le^{3}\Ti
\end{equation}
independently of whether $\dim(\yg)$ equals $\Li^{2}$ or $\Ti^{2}$. This
convention has several advantages, and implies that the hypervolume of a
four-dimensional region, given by the integral of $\ygv$, also has
dimension $\Li^{3}\Ti$, see \eqn~\eqref{eq:integration} -- which is a
reasonable result for a
\emph{space\textnormal{($\Le^{3}$)}time\textnormal{($\Ti$)}} region.

In general the metric $\yg$ induces volume, area, and line elements on
three-, two-, and one-dimensional regions. It is convenient to multiply
these elements by appropriate powers of $c$ so that the region's volume has
intuitive dimensions, such as $\Li^{3}$ for a spacelike three-dimensional
region and $\Li\Ti$ for a timelike two-dimensional one. Indeed the
definition of proper time~\eqref{eq:proper_time_c} does exactly this,
including a factor $1/c$ in the induced line element on a timelike curve.

The \emph{inverse proper volume element} is the 4-vector field $\ygv^{-1}$,
with twisted orientation, having unit generalized inner product with the
proper volume element: $\ygv^{-1} \ii \ygv = 1$. Its intrinsic dimension is
therefore
\begin{equation}
  \label{eq:dim_inv_volume_elem}
  \dim\bigl(\ygv^{-1}{}\i{^{\mul{\q\q\q\q}}}\bigr) =
  \dim(\ygv)^{-1} \equiv \Le^{-3}\Ti^{-1}
\end{equation}
again independently of whether $\dim(\yg)$ equals $\Li^{2}$ or $\Ti^{2}$.
Note that the inverse proper volume element is dimensionally and
numerically different from the tensor  obtained by raising all
indices of $\ygv$.%  It can be proven, for example using an orthonormal
% basis, that the relation between the two is
% \begin{equation}
%   \label{eq:relation_volumes}
%   \ygu = -\frac{1}{c^{2}} \ygv^{-1} \ .
% \end{equation}
% More generally the negative factor is replaced with $(-1)^{s}$, where $s$ is
% the number of negative eigenvalues of the metric tensor.

\medskip

The proper volume element appears in the various definitions of the
\emph{star operator}
\citep[\eg][\sect~V.A.4]{choquetbruhatetal1977_r1996}[Box~4.3]{misneretal1970_r1973}[\sect~IV.24]{burke1985_r1987}
on covectors and forms. This operator usually acts by first rising all
indices of a covector and then taking the generalized inner product (see
\sect~\ref{sec:tensor_ops}) with the proper volume element. For example,
for a 2-covector $\omega \equiv \omega\i{_{\mul{\q\q}}}$
\begin{equation}
  \label{eq:star_op}
  *\omega \defd (\yg^{-1}\omega\yg^{-1})\ii\ygv \ ,
\end{equation}
with $\yg^{-1}$ appearing twice in this specific case. From the general
definition it's clear that the star operator's effect on the dimension
depends on the degree of the form it operates on. I personally prefer to
avoid the star operator and to explicitly use the inner product with the
proper volume element \citep[\cf][\sects~4.1--2]{bossavit1991}.

\subsection{Four-velocity and projector onto it}
\label{sec:g_4velocity}

The worldline of an observer or of a small body is a timelike curve
$C\colon \ytp \mapsto P(\ytp)$ into spacetime, parameterized by the proper
time $\ytp$. If we assume $\dim(\ytp)=\Ti$, then according to the
discussions in \sects~\ref{sec:curves}, \ref{sec:g_intr_dim},
\ref{sec:g_inv_vol} the condition that the curve's parameter should beat
the proper time leads to two different normalization conditions for the
\emph{4-velocity tangent vector} $\yU \equiv \yU\i{^{\q}} \defd \yC$,
depending on the choice of dimension for the metric:
\begin{equation}
  \label{eq:proper_time_condition}
  \begin{aligned}
   \frac{1}{c^{2}}\, \yU(\tau)\,\yg\,\yU(\tau) &= \pm 1
    &&\text{if }\dim(\yg) \defd \Le^{2} \ ,
      \\
    \yU(\tau)\,\yg\,\yU(\tau) &= \pm 1
    &&\text{if }\dim(\yg) \defd \Ti^{2} \ ,
  \end{aligned}
\end{equation}
where $\pm 1$ is the sign of the time-time component of the metric.
Either equation is dimensionless under its specific condition.
Independently of normalization conditions the 4-velocity has intrinsic
dimension $\dim(\yU\i{^{\q}}) = \Ti^{-1}$.

Either condition leads to the same expression for the 4-velocity in a system of rectangular Cartesian coordinates $(t,x,y,z)$ with $\dim(t,x,y,z) = (\Ti,\Le,\Le,\Le)$, adapted to an inertial observer:
\begin{equation}
  \label{eq:4-velocity_coords}
  \yU  = \yl \de_{t} + \yl V^{r} \de_{x^{r}} \ ,
  \qquad  \yl \defd 1/\sqrt{1- \yV^{2}/c^{2}}
\end{equation}
where $\yl$ is the (dimensionless) Lorentz contraction factor and $V^{r}$,
$r \in \set{1,2,3}$, are the components of the \emph{coordinate} 3-velocity
$\yV$.

Care must be taken with the covariant 4-velocity $\yUu$ usually obtained by lowering the index of $\yU$. Under the choice $\dim(\yg) \defd \Le^{2}$, simply lowering the index would lead to an object with the peculiar dimensions $\Le^{2}\Ti^{-1}$. This suggests that some $c$ factors should be included in its definition. Possibly some $\pm$ sign should also be included, depending on the sign of the time-time component of the metric, if we want $\yU \otimes \yUu$ to act as an \emph{idempotent} projector onto the proper-time axis in $3+1$ formulations of general relativity.

I find it convenient to define the covariant 4-velocity as
\begin{equation}
  \label{eq:covariant_4velocity}
  \yUu \defd
  \frac{\yU \yg}{\yU(\tau)\,\yg\,\yU(\tau)}
% \equiv \begin{cases}
%     \pm\dfrac{1}{c^{2}}\,\yU \yg & \text{if }\dim(\yg)\defd \Le^{2}
%     \\[\jot]
%     \pm\yU \yg & \text{if }\dim(\yg)\defd \Ti^{2}
%   \end{cases}
 \ .
\end{equation}
%where $\pm 1$ is again the sign of the time-time component of the metric.
This definition has the following advantages:
\begin{itemize}
\item it has dimensions $\dim(\yUu) = \Ti$,
\item in rectangular Cartesian inertial coordinates it has the expression
\begin{equation}
  \label{eq:cov_4-velocity_coords}
  \yUu  = \yl \di t - \sum_{r} \tfrac{1}{c^{2}} \yl V^{r} \di x^{r}
\end{equation}
independently of the dimensions and signature of the metric,
\item $\yU \otimes \yUu$ is an idempotent projector onto the time axis,
independently of the dimensions and signature of the metric, and also independently of whether $\yU$ is normalized.
\end{itemize}
% leads to the same expression in rectangular Cartesian
% inertial coordinates, and also to the property $\yU\yUu = -1$, with
% $\dim(\yU)^{-1} = \dim(\yU) = \Ti^{-1}$, independently of the choice of
% dimension for the metric. This property is especially useful in $3+1$
% formulations of general relativity, where $\yU \otimes \yUu$ acts as a
% projector onto the proper-time axis. If you use $\dim(\yg)\defd\Le^{2}$ and
% define the covariant 4-velocity as $\yU\yg$ instead, then keep in mind that
% such a projector requires an extra factor $1/c^{2}$.

\medskip

Alternatively we can stipulate \citep[\eg][\eqn~(5) p.~920]{eckart1940c} that when $\dim(\yg) \defd \Le^{2}$ the proper time is actually a \emph{length}: $\dim(\ytp) \defd \Le$. Then  $\dim(\yUu)^{-1} = \dim(\yU) = \Le^{-1}$, the $c^{2}$ factors in formula~\eqref{eq:proper_time_condition} above disappear, and the special coordinate expressions~\eqref{eq:4-velocity_coords} and \eqref{eq:cov_4-velocity_coords} acquire factors $1/c$ and $c$.

% -- usually but not necessarily the proper
% time, see \sect~\ref{sec:metric} -- with $\dim(t) = \Ti$. The 4-velocity of
% the body at a given spacetime event is the tangent vector to the worldline
% thus parameterized. From \eqn~\eqref{eq:dim_velocity} it follows that
% \emph{the 4-velocity has intrinsic dimension \textnormal{$\Ti^{-1}$}}.

% % You might find this result counter-intuitive. Probably the reason is that
% % you're thinking in terms of components. There is no contradiction: 
% If we use a system of coordinates with dimensions $\Le$ and $\Ti$, then the
% components of the 4-velocity have dimensions $\Le/\Ti$ and $\Un$, in accord
% with intuition. The fact that the intrinsic dimension is $\Ti^{-1}$, with
% no lengths involved, is quite sensible. We have equipped the body with some
% kind of clock, so we can say how much time has passed between two spacetime
% events intersected by the body's worldline. But we cannot say what their
% distance is, because no metric is involved in the definition.

% If we introduce a metric tensor, The \emph{norm} of the 4-velocity,
% calculated using the metric tensor, has dimension $\Le/\Ti$ or $\Un$
% depending on the dimension chosen for the metric tensor, discussed in
% \sect~\ref{sec:metric}.

\subsection{Induced connection and Einstein tensor}
\label{sec:g_einst}

The formulae for the covariant derivative~\eqref{eq:dim_cov_der_vect},
connection coefficients~\eqref{eq:dim_christoffel}, and curvature
tensors~\eqref{eq:dim_torsion_riemann_ricci} remain valid for a connection
compatible with the metric. In this case the connection coefficients can be
obtained from the metric by the formulae
\citep[\sect~V.B.2]{choquetbruhatetal1977_r1996}
\begin{equation}
  \label{eq:christoffel_g}
  \yGa\i{^{i}_{jk}} = \frac{1}{2}
  \biggl(\dex[k]g_{jl} + \dex[j]g_{kl} - \dex[l]g_{jk}\biggr)\,g^{li} \ ,
\end{equation}
and it's easily verified that the dimensions of these coefficients given in
\eqn~\eqref{eq:dim_christoffel} still hold, as do the results for the
curvature tensors~\eqref{eq:dim_torsion_riemann_ricci}.

\medskip

The \emph{scalar curvature} $\ysc$ and the co-co-variant \emph{Einstein
  tensor} $\yG \equiv \yG\i{_{\q\q}}$
\begin{equation}
  \label{eq:curv_einst}
  \ysc \defd %\tr\yRi\i{_{\q}^{\q}}\equiv
  \tr(\yRi\ \yg^{-1})
  \ ,
  \qquad
  \yG\i{_{\q\q}} \defd \yRi -%\i{_{\q\q}} -
  \tfrac{1}{2} \ysc\ \yg %\i{_{\q}^{\q}} 
\end{equation}
have  dimensions
\begin{gather}
  \label{eq:dim_scalcurv}
  \dim(\ysc) = \dim(\yg)^{-1} \equiv
  \begin{cases}
    \Le^{-2}   \quad\text{if }\dim(\yg)\defd      \Le^2 \\
    \Ti^{-2}\quad\text{if }\dim(\yg)\defd      \Ti^2 \\
  \end{cases} \ ,
  \\[\jot]
  \label{eq:dim_einst}
  \dim(\yG\i{_{\q\q}}) = \Un \ ,
\end{gather}
that is, \emph{the twice covariant Einstein tensor is dimensionless,
  independently of the dimension of the metric tensor}.

\section{Stress-energy-momentum tensor}
\label{sec:stressenergy}

Also in the case of the stress-energy-momentum tensor the literature offers
two main choices of intrinsic dimension, independent of the choices for the
metric tensor discussed in the previous section. Moreover, there seems to
be no consensus yet on what the primitive variance type of the
stress-energy-momentum tensor should be. Its operational meaning is still
surrounded by some mystery. Let's try to find its dimension and variance
type through a heuristic approach, which will also show the usefulness of
intrinsic dimensional analysis on differential manifolds.

The stress-energy-momentum tensor for a material continuum at a spacetime
event embodies the volumic energy (comprising internal, kinetic, and rest
energy) $\en$, areic energy flux $\yq_{r}$ (comprising convected volumic
energy and heating), volumic momentum $\yp_{r}$, and stress $\yt_{sr}$
(considered as compressive rather than tensile, and including convected
volumic momentum) of the material at that event.\citep[For the
\enquote{volumic} and \enquote{areic} terminology see][\sect~A.6]{iso2009}
Here the vertical position of the indices $r,s \in \set{x,y,z}$ does
\emph{not} denote any variance type. These quantities are measured by an
inertial observer at that event, using a system of one timelike and three
spacelike coordinates $(t,x,y,z)$. If these coordinates have dimensions
$(\Ti,\Le,\Le,\Le)$, then the dimensions of the quantities are
\begin{equation}\label{eq:dims_stress_components}
  \begin{aligned}
    \dim(\en) &=\Ma\Le^{-1}\Ti^{-2} \equiv  \En\Le^{-3} \ ,
    &
    % \label{eq:dim_yq}
    \dim(\yq_{r}) &=\Ma\Ti^{-3} \equiv  \En\Le^{-2}\Ti^{-1} \ ,
    \\
    \dim(\yp_{r}) &=\Ma\Le^{-2}\Ti^{-1} \equiv  \En\Le^{-4}\Ti \ ,
    &
    % \label{eq:dim_stress}
    \dim(\yt_{rs}) &=\Ma\Le^{-1}\Ti^{-2} \equiv  \En\Le^{-3} \ .
  \end{aligned}
\end{equation}

Suppose we want to construct a tensor $\yT$ having these 16 independent
quantities as components. What should its variance type and its intrinsic
dimension be? I am not assuming the symmetry of this tensor as an a-priori
kinematic property, leaving it instead as a dynamical law enforced by the
Einstein equations; in fact this symmetry only needs to hold for the sum of
the stress-energy-momentum tensors from all kinds of matter.

Since we have 16 components, this tensor should belong to the tensor
product of two tangent spaces, each spanned by four basis elements. There
are four such spaces: vectors, covectors, 3-vectors, and 3-covectors. Let's
use shorthands such as
$\detzy \defd \de_{t} \land \de_{z} \land \de_{y}$ and
$\ditzy \defd \di t \land \di z \land \di y$. These four spaces then have
the following coordinate-induced bases and corresponding dimensions:
\begin{subequations}\label{eq:bases}
  \begin{align}
    \label{eq:base_vec}
    &  (\de_{t}, \de_{x}, \de_{y}, \de_{z})\ \text{:}
    &&\text{\small }\ (\Ti^{-1}, \Le^{-1}, \Le^{-1}, \Le^{-1}) \ ,
    \\
    \label{eq:base_covec}
    &  (\di t, \di x, \di y, \di z) \ \text{:}
    &&\text{\small }\ (\Ti, \Le, \Le, \Le) \ ,
    \\
    \label{eq:base_trivec}
    &  (\dexyz, \detzy, \detxz, \detyx)\ \text{:}
    &&\text{\small }\  (\Le^{-3}, \Le^{-2}\Ti^{-1}, \Le^{-2}\Ti^{-1}, \Le^{-2}\Ti^{-1}) \ ,
    \\
    \label{eq:base_tricovec}
    &  (\dixyz, \ditzy, \ditxz, \dityx)\ \text{:}
    &&\text{\small }\  (\Le^{3}, \Le^{2}\Ti, \Le^{2}\Ti, \Le^{2}\Ti) \ ,
  \end{align}
\end{subequations}
where the orderings are chosen to minimize the minus signs appearing from
inner products with a volume element. There are therefore $4 \times 4$
possible tensor-product spaces, each constructed by the product of two of
the four spaces above; % , in all possible combinations;
and thus sixteen possible alternatives to represent our
stress-energy-momentum tensor. Volumic energy is intuitively associated
with the purely timelike component of this tensor, stress with the purely
spacelike components, and areic energy flux and volumic momentum with the
mixed timelike-spacelike components.

Consider the following first alternative, obtained from the tensor product
of the space~\eqref{eq:base_vec} with itself; omit $y$- and $z$-terms for
brevity:
\begin{equation*}
\yT \stackrel{?}{=}    \en\  \de_{t}\otimes\de_{t} +
  \yq_{x}\  \de_{t}\otimes\de_{x} +
  \yp_{x}\  \de_{x}\otimes\de_{t} +
  \yt_{x x}\  \de_{x}\otimes\de_{x} + \dotsb
\end{equation*}
% \begin{equation*}
%   \begin{aligned}
%     \yT \stackrel{?}{=}
%   &\ \en\  \di t\otimes\di x +
%   \yq_{x}\  \di t\otimes\di x +
%   \yp_{x}\  \di x\otimes\di t +
%   \yt_{xx}\  \di x\otimes\di x + \dotsb
% \\ %   \\[-\jot]\text{\footnotesize or}&\\[-\jot]
% \yT \stackrel{?}{=}    &\ \en\  \de_{t}\otimes\de_{t} +
%   \yq_{x}\  \de_{t}\otimes\de_{x} +
%   \yp_{x}\  \de_{x}\otimes\de_{t} +
%   \yt_{x x}\  \de_{x}\otimes\de_{x} + \dotsb
% \\ %   \\[-\jot]\text{\footnotesize or}&\\[-\jot]
%  \yT \stackrel{?}{=} &\ \en\  \de_{t}\otimes\di t +
%   \yq_{x}\  \de_{t}\otimes\di x +
%   \yp_{x}\  \de_{x}\otimes\di t +
%   \yt_{x x}\  \de_{x}\otimes\di x + \dotsb
% \\ %  \\[-\jot]\text{\footnotesize or}&\\[-\jot]
% \yT \stackrel{?}{=}  &\ \en\  \di t\otimes\de_{t} +
%   \yq_{x}\  \di t\otimes\de_{x} +
%   \yp_{x}\  \di x\otimes\de_{t} +
%   \yt_{x x}\  \di x\otimes\de_{x} + \dotsb
%   \end{aligned}
% \end{equation*}
The first and third summands of this expression have incompatible intrinsic
dimensions $\En\Le^{-3}\Ti^{-2}$ and $\En\Le^{-5}$. This alternative is
therefore rejected because dimensionally inconsistent. Similar dimensional
analyses on the remaining fifteen alternatives show that only four are
dimensionally consistent:
\begin{subequations}
  \label{eq:T_alternatives}
  \begin{align}
    \label{eq:T_co_contra}
    &\yT\i{_{\q}^{\q}} = -\en\  \di t\otimes\de_{t} 
      -\yq_{x}\  \di t\otimes\de_{x} +
      \yp_{x}\  \di x\otimes\de_{t} +
      \yt_{x x}\  \di x\otimes\de_{x} + \dotsb
    \\[\jot]
    \label{eq:T_co_co3}
    &\!\begin{multlined}[b][0.85\textwidth]
      \yT\i{_{\q\mul{\q\q\q}}} =
      -\en\  \di t\otimes\dixyz 
      -\yq_{x}\  \di t\otimes\ditzy +{}\\
      \yp_{x}\  \di x\otimes\dixyz 
      +\yt_{x x}\  \di x\otimes\ditzy + \dotsb
    \end{multlined}
    \\[\jot]
    \label{eq:T_contra3_contra}
    &\!\begin{multlined}[b][0.85\textwidth]
      \yT\i{^{\mul{\q\q\q}\q}} = -\en\  \dexyz \otimes\de_{t} 
      -\yq_{x}\  \dexyz\otimes\de_{x} +{}\\
      \yp_{x}\  \detzy \otimes\de_{t} +
      \yt_{x x}\  \detzy\otimes\de_{x} + \dotsb
    \end{multlined}
    \\[\jot]
    \label{eq:T_contra3_co3}
    &\!\begin{multlined}[b][0.85\textwidth]
      \yT\i{^{\mul{\q\q\q}}_{\mul{\q\q\q}}} = -\en\  \dexyz \otimes\dixyz 
      -\yq_{x}\  \dexyz\otimes\ditzy +{}\\
      \yp_{x}\  \detzy \otimes\dixyz 
      + \yt_{x x}\  \detzy\otimes\ditzy
      + \dotsb
    \end{multlined}
  \end{align}
\end{subequations}
where the particular signs of the components, which are not determined by
dimensional analysis, will be motivated later. Our analysis of the
intrinsic dimensions therefore restricts the stress-energy-momentum tensor
to be one of the four alternatives above, with their kinds of orientation,
straight or twisted, still undetermined. Note that if we had conflated time
and length dimensions in \eqns~\eqref{eq:dims_stress_components} and
\eqref{eq:bases} by introducing $c$ factors, then dimensional analysis
wouldn't have led to any restrictions: all sixteen alternatives would have
been dimensionally consistent. This shows the usefulness -- at times at
least -- to keep dimensions well distinct.

To further restrict the possibilities let's consider three additional and
interrelated heuristic arguments.

First, the notions of volumic energy and momentum, areic energy flux, and
stress imply some kind of integration over three-dimensional spacelike or
timelike regions. Such integration needs a 3-form and thus excludes
alternatives~\eqref{eq:T_co_contra} and \eqref{eq:T_contra3_contra}.

Second, the total energy measured within a \emph{topologically} specified three-dimensional spatial region of an observer's orthogonal hyperplane is considered to be independent of the \enquote{volume} of that region, whether this volume be reckoned by a physical metric or by arbitrary coordinate intervals. % (just like
% charge, although the latter is also independent of the observer's
% motion). 
% \citep[\cf][\sect~4.28]{landauetal1939_t1996}
The \emph{volumic} energy therefore \emph{does} depend on the volume of the
region and must change accordingly. Similar arguments hold for the surface
energy flux. Only the second tensor alternative~\eqref{eq:T_co_co3} above
is consistent with these requirements.

This volume-scaling argument also suggests that the volumic-momentum part $\yp_{x}\ \di x\otimes\dixyz$ should not be interpreted as an areic mass flux $\Ma/(\Le^{2}\Ti)$, that is, as something that needs to be integrated over a surface and over time to yield a mass. The second factor of the basis element $\di x\otimes\dixyz$ indicates that integration should instead happen over a volume, to yield a momentum: % We
% are thus forced to interpret the momentum-density part in terms of an
% \emph{action} density per length instead:
$\dim(\yp_{r}) = (\Ma\Le\Ti^{-1})/\Le^{3}$.
% the $\Le^{-2}$ scaling in the component
% $\yp_{x}$ and the $\Le^{4}$ scaling in the corresponding basis element
% $\di x\otimes\dixyz$ don't match. They do match, however, if we interpret
% the momentum-density part in terms of energy or action:
% $\Ma\Le^{-2}\Ti^{-1}=\En\Ti\Le^{-4}$.
A similar situation occurs for the stress part, which should be interpreted
as an areic momentum flux:
$\dim(\yt_{rs}) = (\Ma\Le\Ti^{-1})/(\Le^{2}\Ti)$. It is worth noting
that scaling reinterpretations of this kind occur, even for more
components, in all other fifteen alternatives for the
stress-energy-momentum tensor. Intrinsic dimensional analysis alone thus
suggests that there is a difference between mass flux and
momentum, a fact that relativity theory makes quite clear \citep[see \eg]{eckart1940c}.% , energy, action, and seems to give the latter quantity a more
% fundamental status.

Third, the value of the energy density should not change under a change in
the orientation of the spacelike coordinates: the total energy in a small region of space remains the same if we decide to replace the coordinate $x$ with $-x$. The 3-covector slot in
alternative~\eqref{eq:T_co_co3} should therefore have a twisted
orientation. This means that the 3-form $\dixyz$ actually has an inner
orientation in the positive $t$ direction, the 3-form $\ditzy$ in the
positive $x$ direction, and so on \citep[To visualize this \cf][Fig.~6 and
table in \sect~II.8 p.~31]{schouten1951_r1989}[and][Fig.~22.10]{burke1985_r1987}.
% \begin{equation*}
%   \en\  \di t\otimes\dixyz =
%   \en'\  \di t\otimes\di x'y'z' \ ,
% \end{equation*}
% To further restrict the possibilities let's consider changes of scale and
% orientation. The total energy measured by an observer within a
% topologically given, spacelike three-dimensional volume element is
% considered to be independent of the spacelike metric properties around that
% element . The energy density therefore will depend on the
% spakelike metric properties and change accordingly.
% For example, if we switch from metres
% to decimetres, the new coordinates $(x',y',z')$ are obtained multiplying
% the old by $10\,\textrm{dm}/\textrm{m}$, and the new energy density $\en'$
% dividing the old by $(10\,\textrm{dm}/\textrm{m})^{3}$. 

\medskip

A heuristic application of intrinsic dimensional analysis combined with
integration, scaling, and orientation arguments thus tells us that the
stress-energy-momentum tensor has variance type $\yT\i{_{\q\mul{\q\q\q}}}$,
that is, it's a covector-valued 3-covector, or a four-times-covariant
tensor completely antisymmetric in three slots. The 3-covector part has a
twisted orientation. This tensor has the dimension of an \emph{action}:
\begin{equation}
    \label{eq:final_dim_T}
  \dim\bigl(\yT\i{_{\q \mul{\q\q\q}}}\bigr) = \En\Ti \equiv \Ma\Le^{2}\Ti^{-1} \ .
\end{equation}
This result agrees with the stress-energy-momentum tensor that appears for example in Einstein's original work \citep[\sect~C.9, discussion before \eqn~(42a)]{einstein1914b}, other early works \citep[\eg][\sect~IV.54]{pauli1921_t1958}[\sect~13]{cartan1923}[\sect~7]{brillouin1924}, and more recent works \citep[\sect~VIII.3]{fokker1960_t1965}[\chap~14 Exercise~14.18, \chap~15, \sect~21.3]{misneretal1970_r1973}[\sect~7 Table~I]{post1982c}{hehletal1986,gotayetal1992,gronwaldetal1997,castrillonlopezetal2008,castrillonlopezetal2009}[see also][for similar conclusions in general manifolds and in Newtonian mechanics]{segevetal1999,kansoetal2007}; and also in Truesdell \amp\ Toupin \citep[\sect~F.IV.288]{truesdelletal1960}, who try to find an expression universally valid in Newtonian, Lorentzian, and general-relativistic mechanics. The commonly encountered versions of this tensor with only two slots are discussed below.
% \citep[\enquote{$^{\displaystyle *}\mathte{T}$}]
Note that some of these works use a once covariant and once contravariant
\enquote{V(olume)-tensor} or \enquote{tensor density}, which has variance
type $\yT\i{_{\q}^{\q}_{\mul{\q\q\q\q}}}\,$. Such an object, however, is
geometrically equivalent to the variance type $\yT\i{_{\q \mul{\q\q\q}}}\,$;
their independent components have the same transformation law under changes
of coordinates \citep[\sect~II.8 p.~30]{schouten1951_r1989} (this is why I
chose a calligraphic letter to denote this tensor). % They are related by
% an inner product with $\omega\otimes\omega^{-1}$, where $\omega$ is an
% arbitrary non-vanishing 4-form; this last object has only one independent
% component, which transforms as the scalar $1$.

The signs of the components of $\yT$ depend on the signature of the metric
$\yg$. If the latter has signature $(-,+,+,+)$, then the energy components
have negative sign, as in \eqn~\eqref{eq:T_co_co3}. If the metric has
opposite sign, that is, signature $(+,-,-,-)$, then $\yT$ has opposite sign
to \eqn~\eqref{eq:T_co_co3} as well, and its momentum components are
negative instead.

\medskip

The literature cited above arrive at this kind of stress-energy-momentum
tensor through inductive generalization, often via electromagnetic theory,
of the stress tensor of Newtonian mechanics; or from principles of virtual
work; or from variational principles with an action Lagrangean \citep[\cf\
also][]{hilbert1915,hilbert1917}[\sect~3.3]{hawkingetal1973_r1994}, from
which it easily follows that this tensor should have the intrinsic
dimension of an action; or from combinations of these approaches. The
operational meaning of this tensor is therefore still unclear in my
opinion.

In the presence of a metric tensor we can of course obtain stress-energy-momentum tensors of different variance types by means of inner products with the proper volume element and its inverse, and by raising and lowering indices. But the question of the operational meaning and primitive variance type of this tensor are important, for example, in field theories not based on a metric, or for the formulation of constitutive equations \citep[\cf][\chap~G]{truesdelletal1960}{marsdenetal1983b_r1994,gotayetal1992}. Extensive investigations were made by Gotay \etal\ \citep{gotayetal1998,gotayetal2004,gotayetal2006}; and by Segev \citep{segev2002}[see also][]{segev2000b}, who interprets the stress-energy-momentum tensor as a linear map from the four-dimensional flux of a conserved quantity, such as charge or baryonic number, to the flux of energy. Since such fluxes are represented by 3-forms, he arrives at the fourth alternative~\eqref{eq:T_contra3_co3} above: the 3-vector part of $\yT\i{^{\mul{\q\q\q}}_{\mul{\q\q\q}}}$ can be contracted with a 3-form, yielding another 3-form. This interesting interpretation doesn't seem to work out dimensionally, however. For example, the intrinsic dimension of the four-dimensional charge flux is charge itself, $\Ch \equiv \Cu\Ti$; in order to yield an energy flux, which has intrinsic dimension of energy $\En$, the tensor $\yT\i{^{\mul{\q\q\q}}_{\mul{\q\q\q}}}$ should then have intrinsic dimension $\En\Ch^{-1}$ according to the results of \sect~\ref{sec:tensor_ops}. The dimension of charge would then have to appear in Einstein's constant $\yk$ (see \sect~\ref{sec:einstein_eq}), because it cannot be eliminated by using the metric tensor or the proper volume element to obtain alternative variance types. Similar problems occur with the flux of baryonic number, which has dimension of amount of substance $\textsf{N}$.

\medskip

Some authors
\citep[\eg][\sect~V.55]{fock1955_t1964}[\sect~4.1]{mcvittie1956_r1965}[\sect~10.1]{adleretal1965_r1975}
conceive the stress-energy-momentum tensor in terms of mass rather than
energy (Fock \citep[\sect~II.31]{fock1955_t1964} calls it the \enquote{mass
  tensor}), and therefore assign to its covector-valued 3-covector form
$\yT\i{_{\q \mul{\q\q\q}}}$ the dimension of mass-time, that is, an action
divided by squared velocity:
\begin{equation}
  \label{eq:final_dim_T_MT}
  \dim\bigl(\yT\i{_{\q \mul{\q\q\q}}}\bigr) = \Ma\Ti \equiv \En\Le^{-2}\Ti^{3} \ .
\end{equation}
With this intrinsic dimension, however, not all components of the
stress-energy-momentum tensor have intuitive meanings and dimensions when
coordinates with dimensions $(\Ti,\Le,\Le,\Le)$ are used. The two choices
\eqref{eq:final_dim_T}, \eqref{eq:final_dim_T_MT} differ by a factor
$c^{2}$.

\medskip

The two dimensional choices for the metric, \eqns~\eqref{eq:dim_g} and
\eqref{eq:dim_g_c}, and for the stress-energy-momentum tensor,
\eqns~\eqref{eq:final_dim_T} and \eqref{eq:final_dim_T_MT}, appear in all
four combinations in the literature. For example, $\dim(\yg)=\Ti^{2}$ and
$\dim(\yT)=\En\Ti$ is used by Synge \citep[\sects~IV.4--5]{synge1960b};
$\dim(\yg)=\Le^{2}$ and $\dim(\yT)=\Ma\Ti$ is used by Fock
\citep[\sects~V.54--55]{fock1955_t1964} and Adler \etal\
\citep[\sect~10.1]{adleretal1965_r1975}; $\dim(\yg)=\Ti^{2}$ and
$\dim(\yT)=\Ma\Ti$ is used by McVittie
\citep[\sect~4.1]{mcvittie1956_r1965} and possibly Kilmister
\citep[\chaps~II--III; he seems to shift to natural units at some
point]{kilmister1973}. Most other works use $\dim(\yg)=\Le^{2}$ and
$\dim(\yT)=\En\Ti$. These combinations lead to three possible values for
Einstein's constant, discussed in the next section.

\medskip

To obtain a twice covariant tensor to be used in the Einstein equations, we
first take the inner product of the inverse proper volume element with the
3-covector (that is, antisymmetric) part of $\yT\i{_{\q \mul{\q\q\q}}}$,
obtaining a co-contravariant tensor. % $\yTc \equiv
% \yTc{}\i{_{\q}^{\q}}$
% \begin{equation}
%   \label{eq:T-co_contra}
%   \yTc \defd \yT\i{_{\q\mul{\q\q\q}}} \rii \ygv^{-1} \ .
% \end{equation}
% In general coordinates, and if $\dim(\yg) \defd \Le^{2}$, this operation
% corresponds to dividing all components of $\yT$ by
% $\frac{1}{c}\sqrt{\smash[b]{\abs{\det(g\i{_{ij}})}}}$, according to
% \eqn~\eqref{eq:volume_elem}, and so can be interpreted as passing from a
% tensor density to an ordinary tensor. Roughly speaking, whereas the tensor
% $\yT$ reports the three-dimensional spacelike or timelike densities with
% respect to \emph{coordinate} volume, the tensor $\yTc$ reports them with
% respect to physical volume determined by the metric. This interpretation
% must be taken with a pinch of salt because this tensor 
Then we lower the new contravariant slot by means of the metric tensor. The
combined operation yields
\begin{gather}
  \label{eq:co_co_T}
  \yTe\i{_{\q\q}} \defd
  \bigl(\yT\i{_{\q\mul{\q\q\q}}} \rii \ygv^{-1} \bigr) \yg \ .
\end{gather}

According to this definition and \eqns~\eqref{eq:final_dim_T_MT},
\eqref{eq:dim_g}, \eqref{eq:dim_g_c}, \eqref{eq:dim_volume_elem}, the
co-co-variant tensor $\yTe$ has three possible intrinsic dimensions,
depending on the choices of dimensions of $\yg$ and $\yT$:
\begin{multline}
\label{eq:dim_co_co_T}
\dim(\yTe\i{_{\q\q}}) =
\dim\bigl(\yT\i{_{\q \mul{\q\q\q}}}\bigr) \dim(\yg) \dim(\ygv)^{-1}
  ={}\\
  \begin{cases}
    \Ma\Le\Ti^{-2} \equiv \En\Le^{-1} & \quad\text{if}\ 
\dim(\yg)\defd \Le^2 \ ,\ \dim(\yT)\defd \En\Ti
    \\[\jot]
\Ma\Le^{-1} \equiv \En\Le^{-3}\Ti^{2} & \quad\text{if}\ 
 \begin{cases}
\dim(\yg)\defd \Ti^2 \ ,\ \dim(\yT)\defd \En\Ti 
    \quad\text{or}\\
\dim(\yg)\defd \Le^2 \ ,\ \dim(\yT)\defd \Ma\Ti 
 \end{cases}
   \\[2\jot]
    \Ma\Le^{-3}\Ti^{2} \equiv \En\Le^{-5}\Ti^{4} & \quad\text{if}\ 
\dim(\yg)\defd \Ti^2 \ ,\ \dim(\yT)\defd \Ma\Ti
  \end{cases}
\end{multline}
All three possibilities, which differ by factors $c^{2}$, appear in the
literature: see the works cited after \eqn~\eqref{eq:final_dim_T}
concerning the combinations of dimensions for metric and
stress-energy-momentum.

It may be useful to write the coordinate expressions of the tensors $\yTe\i{_{\q\q}}$ and $\yTu\i{^{\q\q}}$, obtained from $\yT\i{_{\q\mul{\q\q\q}}}$, in the case with $\dim(\yg)\defd \Le^2$, $\dim(\yT)\defd \En\Ti$, and coordinate system with diagonal metric $\yg = \pm(-c^{2} \di t \otimes \di t + \di x \otimes \di x + \dotsb)$, commonly encountered in the literature:
\begin{align}
      \mbox{}\hspace{-1ex}\yTe\i{_{\q\q}} &=
      c^{2}\en\  \di t\otimes\di t 
      -\yq_{x}\  \di t\otimes\di x
      -c^{2}\yp_{x}\  \di x\otimes\di t 
                        +\yt_{x x}\  \di x\otimes\di x + \dotsb
  \label{eq:common_T_coco}
  \\
      \mbox{}\hspace{-1ex}\yTu\i{^{\q\q}} &=
      \frac{1}{c^{2}}\en\  \de_{t}\otimes\de_{t} 
      +\frac{1}{c^{2}}\yq_{x}\  \de_{t}\otimes\de_{x} 
      +\yp_{x}\  \de_{x}\otimes\de_{t} 
      +\yt_{x x}\  \de_{x}\otimes\de_{x} + \dotsb
    \label{eq:common_T_covcov}
\end{align}
where the $c$ factors can be freely interpreted as part either of the
components or of the $t$ coordinate. Remember that the stress is here
considered as compressive rather than tensile, and that the Einstein
equations require in particular that $\yq_{x} = c^{2} \yp_{x}$ and so on
(heat flux carries momentum \citep[p.~923]{eckart1940c}).

\section{The constant in the Einstein equations}
\label{sec:einstein_eq}

We finally arrive at the Einstein equations,
\begin{equation}
  \label{eq:einstein}
  \yG%\i{_{\q}^{\q}}
  = \yk \yTe%\i{_{\q}^{\q}}
\end{equation}
sometimes seen with a minus sign, depending on the signature of the metric
and of alternative definitions of the curvature tensors \citep[see the
\emph{Table of sign conventions} on the final pages
of][]{misneretal1970_r1973}. $\yk$ is Einstein's constant.

% According to our definitions of $\yG\i{_{\q\q}}$,
% \eqn~\eqref{eq:dim_einst}, and $\yTe\i{_{\q\q}}$, \eqn~\eqref{eq:co_co_T},
The equations above are considered in their co-co-variant form. This form
is convenient because the left side is then dimensionless (its intrinsic
dimension is $\Un$), independently of the dimension of the metric tensor,
as explained in \sect~\ref{sec:g_einst}, \eqn~\eqref{eq:dim_einst}. We
therefore find that the equality
\begin{equation}
  \label{eq:dim_k_general}
  \dim(\yk) = \dim(\yTe\i{_{\q\q}})^{-1} 
\end{equation}
must always hold, for all choices of dimensions for the metric and
stress-energy-momentum tensors. Combining this equation with the results
for $\yTe$, \eqn~\eqref{eq:dim_co_co_T}, we find three possible
conventions:
\begin{multline}
\label{eq:dim_k_choices}
\dim(\yk) ={}\\
  \begin{cases}
    \Ma^{-1}\Le^{-1}\Ti^{2} \equiv \En^{-1}\Le
&\quad\text{if}\ \dim(\yg)\defd \Le^2 \ ,\ \dim(\yT)\defd \En\Ti
    \\[\jot]
    \Ma^{-1}\Le \equiv \En^{-1}\Le^{3}\Ti^{-2}
 &\quad\text{if}\ 
 \begin{cases}
\dim(\yg)\defd \Ti^2 \ ,\ \dim(\yT)\defd \En\Ti 
    \quad\text{or}\\
\dim(\yg)\defd \Le^2 \ ,\ \dim(\yT)\defd \Ma\Ti 
 \end{cases}
    \\[2\jot]
    \Ma^{-1}\Le^{3}\Ti^{-2} \equiv \En^{-1}\Le^{5}\Ti^{-4}
     &\quad\text{if}\ 
\dim(\yg)\defd \Ti^2 \ ,\ \dim(\yT)\defd \Ma\Ti
  \end{cases}
\end{multline}

Einstein's constant $\yk$ can therefore be obtained from Newton's
gravitational constant $\dim(G)=\Ma^{-1}\Le^{3}\Ti^{-2}$ (this is not the
Einstein tensor $\yG$!) and the speed of light $\dim(c)=\Le\Ti^{-1}$ only
in the following three ways, with the $8\pu$ factor coming from the
Newtonian limit:
\begin{equation}
\label{eq:dim_k_Gc}
\yk =
\begin{cases}
  8\pu G/c^{4} &\quad\text{if}\ \dim(\yg)\defd \Le^2 \ ,\ \dim(\yT)\defd \En\Ti
 \\[\jot]
  8\pu G/c^{2} &\quad\text{if}\ 
 \begin{cases}
\dim(\yg)\defd \Ti^2 \ ,\ \dim(\yT)\defd \En\Ti 
    \quad\text{or}\\
\dim(\yg)\defd \Le^2 \ ,\ \dim(\yT)\defd \Ma\Ti 
 \end{cases}
\\[2\jot]
  8\pu G  &\quad\text{if}\ 
\dim(\yg)\defd \Ti^2 \ ,\ \dim(\yT)\defd \Ma\Ti
  \end{cases}
\end{equation}
As we saw in the discussion of the literature cited in
\sect~\ref{sec:stressenergy}, the first convention is the most common. The
second convention appears for example in Fock \citep[\sect~55 \eqns~(55.15)
and~(52.06)]{fock1955_t1964} and Adler \etal\ \citep[\sect~10.5
\eqn~(10.98)]{adleretal1965_r1975}. The third, interesting convention would
appear in McVittie if he didn't cheat a factor $1/c^{2}$ into $\yk$ by
writing the Einstein equations \citep[\sect~4.2
\eqn~(4.107)]{mcvittie1956_r1965} as \enquote{\,$\yG = \yk c^{2} \yTe$\,}.

A fourth possibility is discussed by Post \citep[\sect~8]{post1982c}. If we choose a dimensionless metric tensor $\dim(\yg)\defd\Un$, and define the volume element as $\ygv \defd \sqrt{\smash[b]{\abs{\det(g\i{_{ij}})}}}\ \dix[0]\land\dix[1]\land\dix[2]\land\dix[3]$ so that $\dim(\ygv)=1$, then from \eqns~\eqref{eq:dim_co_co_T} and \eqref{eq:dim_k_general} we find that $\dim(\yk)=(\En\Ti)^{-1}$, the dimension of an inverse action. There is no way to obtain such a dimension with a product of powers of $G$ and $c$. If we include powers of Planck's constant $h$, then no powers of $G$ and $c$ can appear in $\yk$. The operational meaning of these particular dimensional choices and constants is yet unclear to me; I invite you to read Post's paper.

\section{Summary and conclusions}
\label{sec:summary}

We have seen that dimensional analysis, with its familiar rules, can be
seamlessly performed in Lorentzian and general relativity and in
differential geometry if we adopt the coordinate-free approach typical of
modern texts. In this approach each tensor has an \emph{intrinsic}
dimension, a notion introduced by Schouten and Dorgelo. This dimension
doesn't depend on the dimensions of the coordinates, and is determined by
the physical and operational meaning of the tensor. It is therefore
generally more profitable to focus on the intrinsic dimension of a tensor
rather than on the dimensions of its components. The dimension of each
specific component is easily found by
formula~\eqref{eq:dim_component_generic}: it's the product of the intrinsic
dimension by the dimension of the $i$th coordinate function for each
contravariant index $i$, by the inverse of the dimension of the $j$th
coordinate function for each covariant index $j$. Intrinsic dimensional
analysis seems to rest on two main conventions: the tensor product and the
action of covectors on vectors behave analogously to usual multiplication
for the purposes of dimensional analysis. Alternative, equivalent sets of
conventions could perhaps be considered.

We have also seen that intrinsic dimensional analysis can help us determine
or at least constrain the variance type of candidate tensors, as
exemplified with the stress-energy-momentum tensor in
\sect~\ref{sec:stressenergy}. We found or re-derived some essential results
for general relativity, in particular that the Riemann
$\yR\i{^{\q}_{\q\mul{\q\q}}}$, Ricci $\yRi\i{_{\q\q}}$, and twice covariant
Einstein $\yG\i{_{\q\q}}$ curvature tensors are dimensionless. These
results could be of importance for current research involving scales and
conformal factors \citep[\eg][]{roehretal2005,cadonietal2019}. We also
discussed the operational reasons behind two common choices of dimensions
for the metric and stress-energy-momentum tensors.

Since the dimensions of the components are usually different from the
intrinsic dimension and depend on the coordinates, I recommend to avoid
statements such as \enquote{the tensor $A\i{_{i}^{jk}}$ has dimension
  $\Xx$}, which leave it unclear whether \enquote{$A\i{_{i}^{jk}}$} is
meant to represent the tensor in general (as in Penrose \amp\ Rindler's
notation), or to represent its set of components, or to represent only a
specific component.

For the dimensional analysis of tensorial objects in electrodynamics,
which wasn't discussed in this note, see for example Truesdell \amp\ Toupin
\citep[\chap~F]{truesdelletal1960} and Hehl \amp\ Obukhov
\citep{hehletal2003,hehletal2004b}.

% Finally, the above analysis, specific to general relativity, can be made
% along the same lines also for Newtonian mechanics.

\medskip

Dimensional analysis remains a controversial, obscure, but fascinating
subject still today, 60 years from Truesdell \amp\ Toupin's remark quoted
in the Introduction. For an overview of some recent and creative approaches
to it, going beyond Bridgman's text \citep{bridgman1922_r1963} (whose point
of view is in many respects at variance with modern developments: see the
following references), I recommend for example the works by Mari \etal\
\citep{marietal2012,frigerioetal2010}, Domotor and Batitsky
\citep{domotor2017,domotoretal2016,domotor2012}, Kitano \citep{kitano2013},
the extensive analysis by Dybkaer \citep{dybkaer2004_r2010}, the historical
review by de~Boer \citep{deboer1995}, and references therein.

\begin{acknowledgements}
  \ldots to (chronologically) Mariano Cadoni, Ingemar Bengtsson, Iv\'an
  Davidovich, Claudia Battistin for valuable comments on previous drafts.
  To the staff of the NTNU library for their always prompt support. To
  Mari, Miri, Emma for continuous encouragement and affection, and to
  Buster Keaton and Saitama for filling life with awe and inspiration. To
  the developers and maintainers of \LaTeX, Emacs, AUC\TeX, Open Science
  Framework, R, Inkscape, Sci-Hub for making a free and impartial
  scientific exchange possible. %\mbox{}\hfill\autanet

  Note: this paper was rejected by arXiv without any kind of scientific
  justification. %It's sad and alarming that the possibility of pre-print  feedback is curtailed like that.
  
  This work was financially supported initially by the Kavli Foundation and the
  Centre of Excellence scheme of the Research Council of Norway (Roudi
  group), then by the Trond Mohn Research Foundation, grant number BFS2018TMT07.
%\rotatebox{15}{P}\rotatebox{5}{I}\rotatebox{-10}{P}\rotatebox{10}{\reflectbox{P}}\rotatebox{-5}{O}.
%\sourceatright{\autanet}
\end{acknowledgements}

%%%%%%%%%%%%%%%%%%%%%%%%%%%%%%%%%%%%%%%%%%%%%%%%%%%%%%%%%%%%%%%%%%%%%%%%%%%%
%%% Appendices
%%%%%%%%%%%%%%%%%%%%%%%%%%%%%%%%%%%%%%%%%%%%%%%%%%%%%%%%%%%%%%%%%%%%%%%%%%%% 
%\clearpage
% %\renewcommand*{\appendixpagename}{Appendix}
% %\renewcommand*{\appendixname}{Appendix}
% %\appendixpage
% \appendix

%%%%%%%%%%%%%%%%%%%%%%%%%%%%%%%%%%%%%%%%%%%%%%%%%%%%%%%%%%%%%%%%%%%%%%%%%%%%
%%% Bibliography
%%%%%%%%%%%%%%%%%%%%%%%%%%%%%%%%%%%%%%%%%%%%%%%%%%%%%%%%%%%%%%%%%%%%%%%%%%%% 
\renewcommand*{\finalnamedelim}{\addcomma\space}
\defbibnote{prenote}{{\footnotesize (\enquote{de $X$} is listed under D,
    \enquote{van $X$} under V, and so on, regardless of national
    conventions.)\par}}
% \defbibnote{postnote}{\par\medskip\noindent{\footnotesize% Note:
%     \arxivp \mparcp \philscip \biorxivp}}

\printbibliography[prenote=prenote%,postnote=postnote
]

\end{document}